\documentclass[epj]{svjour}
\usepackage{graphics}
\usepackage{amssymb}
\usepackage{amsmath}
\usepackage[pdftex]{graphicx}
\usepackage{epstopdf}
\begin{document}
\title{Finite Volume Cumulant Expansion in QCD-Colorless Plasma}%
\author{M. Ladrem\inst{1,2,3} \and M.A.A. Ahmed\inst{1,3,5} \and Z. Al-Full\inst{1} \and S. Cherif\inst{3,4}%
}                     % Do not remove
%
%\offprints{mmm}          % Insert a name or remove this line
%
\institute{Physics Department, Faculty of Science, Taibah University, Al-Madinah
Al-Munawwarah,Kingdom of Saudia Arabia. \and Physics Department, ENS-Vieux Kouba (Bachir El-Ibrahimi), Algiers, Algeria.
\and \textbf{L}aboratoire de \textbf{P}hysique et de \textbf{M}ath\'ematiques \textbf{A}ppliqu\'ees (LPMA)
,ENS-Vieux Kouba (Bachir El-Ibrahimi),Algiers,Algeria.
\and Sciences and Technologies Department, Ghardaia University, Ghardaia, Algeria.
\and Physics Department, Taiz University in Turba,Taiz, Yemen.}
\date{Received: date / Revised version: date}
\abstract{Due to the finite size effects, the localisation of the phase transition in finite systems and the determination of its order, become an extremely difficult task, even in the simplest known cases.
In order to identify and locate the finite volume transition point $T_{0}(V)$ of the QCD deconfinement phase transition to a Colorless QGP, we have developed a new approach using the finite size cumulant expansion of the order parameter and the $L_{mn}$-method.
The first six cumulants $C_{1,2,3,4,5,6}$ with the corresponding under-normalized ratios(skewness $\Sigma$, kurtosis $\kappa$ ,pentosis $\Pi_{\pm}$ and hexosis $\mathcal{H}_{1,2,3}$) and three unnormalized combinations of them ($\mathcal{O}={\mathcal{\sigma }^{2}
\mathcal{\kappa } }{\mathbf{\Sigma }^{-1} }$, $\mathcal{U} ={\mathcal{\sigma }^{-2} \mathbf{\Sigma }^{-1} }$, $\mathcal{N} = \mathcal{\sigma }^{2} \mathcal{\kappa }$) are calculated and studied as functions of $(T,V)$. A new approach, unifying in a clear and consistent way the definitions of cumulant ratios, is proposed. A numerical FSS analysis of the obtained results has allowed us to locate accurately the finite volume transition point. The extracted transition temperature value $T_{0}(V)$ agrees with that expected $T_{0}^{N}(V)$ from the order parameter and the thermal susceptibility $\chi _{T}\left( T,V\right)$, according to the standard procedure of localization to within about $2\%$. In addition to this, a very good correlation factor is obtained proving the validity of our cumulants method. The agreement of our results with those obtained by means of other models is remarkable.
 % end of PACS codes
} %end of abstract
\maketitle
\section{Introduction}
\label{intro}
\subsection{Phase Transitions and Finite Size Scaling(FSS)}
During the evolution of our beautiful universe from the big-bang instant
until now many phase transitions have occurred at different space-time
scales.For this reason, the physics of phase transitions phenomena
is considered in general to be a subject of great interest to physicists. It is easy to
understand the importance of this subject because firstly, the list of
systems exhibiting interesting phase transitions continues to\ expand,
including the Universe itself, and secondly the theoretical framework of
equilibrium statistical mechanics has found applications in very different
areas of physics like string field theories, cosmology, elementary particle
physics, physics of the chaos, condensed matter ... etc. Phase transitions
occur in nature in a great variety of systems and under a very wide range of
conditions.
\newline
Phase transitions are abrupt changes in the global behavior and in the
qualitative properties of a system when certain parameters pass through
particular values. At the transition point, the system exhibits, by
definition, a singular behavior. As one passes through the transition region the
system moves between analytically distinct parts of the phase diagram.
Depending on which external parameter of interest, there are various measurable quantities which are based on the reaction of a system to its change. We call them Response Functions (RF). If the external parameter corresponds to the temperature, then the response function is called Thermal Response Function (TRF).
Technically, temperature driven phase transitions are characterized by the appearance of
singularities in some TRF, only in the thermodynamic
limit where the volume $V$ and the number of particles $N$ go to infinity,
while the density $\rho={N}/{V}$ remains constant. That is, at the
transition point, some global behavior is not analytic in the infinite
volume limit.
This singularity is according to the standard classification \cite{Hilfer1993} given by the $\mathbf{%
\delta }-$function for a first-order phase transition, while for a
continuous phase transition (second-order)\textbf{,} the singularity has the
form of a power-law function.  We shall frequently refer to the concepts of transition region and transition point in the case of a first order phase transition. By against, in the case of a second order phase transition, we rather use the concept of critical region and critical point.
The singularity in a first order phase transition is entirely due to the phase coexistence phenomenon, for against the divergence in a second-order phase transition is intimately caused by the divergence of the correlation length. Now, if the volume is finite at least in one
dimension with a characteristic size $\ L=V^{1/d}$, the singularity is
smeared out into a peak with finite mathematical properties and Four Finite Size Effects(4FSE)
can be observed \cite{Ladrem2005}:
\newline
(1) the rounding effect of the discontinuities,
\newline
(2) the smearing effect of the singularities,
\newline
(3) the shifting effect of the transition point,
\newline
(4) and the widening effect of the transition region around the transition point.
\newline
These 4FSE have an important consequence putting the first and the second order phase transitions on an
equal footing. The behavior of any physical quantity at the first-order phase
transition is qualitatively similar to that of the second-order phase
transition. However, even in such a situation, it is possible to obtain
information on the critical behavior. Large but finite systems show a
\textit{universal} behavior called \textquotedblleft Finite-Size
Scaling\textquotedblright\ (FSS), allowing to put all the physical systems
undergoing a phase transition in a certain number of universality classes.
The systems in a given universality class display the same critical
behavior, meaning that certain dimensionless quantities have the same values
for all these systems. \textit{Critical exponents} are an example of these
universal quantities. The knowledge of the finite-size dependence of the
various TRF in the vicinity of the phase
transition region provides a very important way to compute, using finite
size scaling extrapolation, the properties of systems in the thermodynamic
limit.
\subsection{Finite Size Effects(FSE) in QCD Deconfinement Phase Transition}
It is well established that Quantum Chromo-Dynamics (QCD) at finite temperature exhibits a typical behavior of a system with a phase transition. At sufficiently high temperatures and/or densities, quarks and gluons are no more confined into hadrons, and strongly interacting matter seems to undergo a
phase transition from hadronic state to what has been called the Quark Gluon
Plasma(QGP) or "Partonic Plasma" (PP). This is a logical consequence of
the parton level of the matter's structure and of the strong
interactions dynamics described by the QCD theory \cite{QCD1998}. The occurrence
of this phase transition is important from a conceptual point of view, as it
implies the existence of a novel state of matter, believed present in the
early universe up to times $\sim 10^{-5}s$. Indeed, the only available experimental way to study this QCD phase transition is to try to create in a laboratory, using ultra-relativistic heavy-ion collisions(URHIC), conditions similar to those in the early moments of the universe, right
after the Big Bang. Due to its similarity to the early universe, an URHIC is often referred to as "little bang". The analysis of the whole results obtained in all experiments at SPS, RHIC and LHC revealed that indeed a new state of matter is formed, consisting of a strongly interacting partons \cite%
{sCQGP}. The existence of this finite volume hot deconfined matter is strongly indicated because some important signatures are observed. One example is the jet quenching phenomenon. According to QCD, high-momentum colored partons produced in the initial stage of a nucleus-nucleus collision will undergo multiple interactions inside the finite volume collision region, generating a parton shower before hadronization. Due to thermal effects the cross section of the hadrons formation and the fragmentation process decrease \cite{NA36,Ladrem2011,Ladrem2013} and to the color confinement property of QCD, only the color singlet part of the quark configurations would manifest themselves as physically observed particles. All hadrons created in the final stage are colorless.  Therefore the whole partonic plasma fireball needs to be in a color singlet state called Colorless QGP (CQGP). For this reason, one can consider the QCD deconfinement phase transition as a transition from local color confinement(d$\sim$1fm) to global color confinement(d$\gg$1fm).
Lattice QCD, a theory formulated on lattice of points in space and time, is an other important framework for investigation of non-perturbative phenomena such as confinement and deconfinement of partons, which are intractable by means of analytic quantum field theories. As is well known, the lattice's space-time volume is finite.
Whereby in both cases of experimental and lattice simulation models, we are dealing with finite systems and, therefore, they require the development of theoretical approaches that can rigorously define the phase transition in a finite volume taking into account the color singlet condition.
Locating the finite volume QCD transition point is a challenge in both theoretical and experimental physics.
\subsection{Motivation}
In the thermodynamic limit there is no problem to locate the transition point since it manifests itself as a singularity point. By cons, in finite volume this singularity is smoothed and is shifted away, consequently the location of the phase transition and the determination of its order become very difficult. The idea of a phase transition is always related to the idea of locating the transition point. Two fundamental questions appear to be very important that we try to answer in the present work. Firstly, how to locate the transition point in finite systems? And secondly, how can we say for sure that a certain physical quantity has a particular behavior when approaching certain point, which may be conceived as the transition point? It is important to have a precise knowledge of the region around the transition point since many quantities of physical interest are just defined in the vicinity of this point. It therefore seems very important to find more sensible quantities to construct new definitions of the finite volume transition point involving a minimum of corrections.
Recently, many works have shown the importance of studying the high order cumulants of thermodynamic fluctuations. For this reason and even in the finite volume case higher-order cumulants and/or generalized ratios of them have been suggested as suitable quantities because they are highly related to the nature of the phase transition and serve as good indicators for a real location of the finite volume transition point.
Mathematically speaking, the thermodynamical fluctuations of any quantity are quantified by cumulants in statistics and are related to generalized ratios of them. Generally, they are defined as derivatives of the logarithm of the partition function with respect to the appropriate chemical potentials.
The cumulant expansion method is then considered by many physicists to be very sensitive to the behavior of the system in the transition region and then is viewed as a promising powerful method to analyse the deconfinement phase transition in finite system \cite{KR2011,Dai2012}. Therefore finding new observables to permit us an accurate localization of the transition point in QCD phase diagram is more than necessary.
From our hadronic probability density function (hpdf) which is related to the total partition function and which contains the whole information about the phase transition as pointed firstly by Gibbs \cite{Gibbs02}, it seems logical to believe that this information survives when the volume of the system becomes finite.
Our basic postulate is that it should be possible to locate the finite volume transition point by defining it as a particular point in each term of the finite size cumulant expansion of the order parameter, suggesting a new approach to solve the problem. We believe that the finite volume cumulant expansion, should show some characteristics as signals of the finite volume transition point. Indeed and in order to identify and locate the finite volume transition point $T_{0}(V)$ of the QCD deconfinement phase transition, we have developed a new approach using the finite size cumulant expansion of the order parameter with the $L_{mn}$-method \cite{Ladrem2005} whose definition has been slightly modified.
The two main outcomes of the present work are: 1) the finite size cumulant expansion of our hpdf gives better estimations, than the Binder cumulant \cite{CLB1986}, for the transition point and even for very small systems. 2) the singularity of the phase transition in thermodynamic limit survives in a clear way even when the volume of the system becomes finite.
\section{Statistical Description of the System Containing the Hadronic Phase
and the Colorless QGP}
\subsection{Exact Colorless Partition Function}
In our previous work, a new method was developed which has allowed us to
accurately calculate physical quantities which describe efficiently the deconfinement
phase transition within the Colorless-MIT bag model using a mixed phase system
evolving in a finite total volume $V$ \cite{Ladrem2005}.The fraction of
volume (defined by the parameter $\mathbf{h}$) occupied by the HG phase is
given by : $V_{HG}=\mathbf{h}V,$ and the remaining volume: $V_{QGP}=(1-%
\mathbf{h)}V$ contains then the CQGP phase. To study the effects of volume
finiteness on the thermal deconfinement phase transition within the QCD
model chosen, we will examine in the following the behavior of some
TRF of the system at a vanishing chemical potential $\left( \mu =0\right) $, considering the two lightest
quarks $u$\ and $d$ $\left( N_{f}=2\right) $, and using the common value $%
B^{1/4}=145MeV$\ for the bag constant.
In the case of a non-interacting phases, the total partition function of the
system can be written as follows:
\begin{equation}
Z_{TOT}(\mathbf{h},V,T,\mu)=Z_{CQGP}(\mathbf{h})Z_{HG}(\mathbf{h})Z_{Vac}(\mathbf{h}%
),
\end{equation}%
where,%
\begin{equation}
Z_{Vac}(\mathbf{h},V,T)=\exp (-(1-\mathbf{h})BV/T),
\end{equation}%
accounts for the confinement of quarks and gluons by the real vacuum
pressure exerted on the perturbative vacuum $\left( B\right) $ of the bag
model. For the HG phase, the partition function is just calculated for a
pionic gas and is simply given by,% {a}_{HG}
\begin{equation}
Z_{HG}(\mathbf{h},V,T)=\exp {{a}_{HG}\mathbf{h}VT^{3}}.
\end{equation}
The exact partition function for a CQGP contained in a volume $V_{QGP},$ at
temperature $T$ and quark chemical potential $\mu$, is determined by:
\begin{equation}
\begin{array}{l}
Z_{CQGP}(T,V_{QGP},\mu ) = \frac{8}{3\pi ^{2}}\int\limits_{-\pi }^{+\pi
}\int\limits_{-\pi }^{+\pi }d\left( \frac{\varphi }{2}\right) d\left(
\frac{\psi }{3}\right) M(\varphi ,\psi ) \\
\times Tr\left[ \exp \left( -\beta \left(
\widehat{H}_{0}-\mu \left( \widehat{N}_{q}-\widehat{N}_{\overline{q}}\right)
\right) +i\varphi \widehat{I}_{3}+i\psi \widehat{Y}_{8}\right) \right],
\end{array}%
\end{equation}
where $M(\varphi ,\psi )$ is the weight function (Haar measure) given by:
\begin{equation}
M(\varphi ,\psi )=\left[ \sin \left( \frac{1}{2}(\psi +\frac{\varphi }{2}%
)\right) \sin (\frac{\varphi }{2})\sin \left( \frac{1}{2}(\psi -\frac{%
\varphi }{2})\right) \right] ^{2},
\end{equation}
$\beta =\frac{1}{T}$ (with the units chosen as: $k_{B}=\hbar =c=1$), and $%
\widehat{H}_{0}$\ is the free quark-gluon Hamiltonian, $\widehat{N}_{q}$ $%
\left( \widehat{N}_{\overline{q}}\right) $\ denotes the (anti-) quark number
operator, and $\widehat{I}_{3}$\ and $\widehat{Y}_{8}$\ are the color
\textquotedblleft isospin\textquotedblright\ and \textquotedblleft
hypercharge\textquotedblright\ operators respectively. Its final expression,
in the massless limit, can be put in the form,
\begin{multline}
Z_{CQGP}(T,V_{QGP},\mu )=\frac{4}{9\pi ^{2}}\int_{-\pi }^{+\pi }\int_{-\pi
}^{+\pi }d\varphi d\psi M(\varphi ,\psi ) \vspace*{0.25cm} \\
\times e^{\mathcal{G}\left(
\varphi ,\psi ,\frac{\mu }{T}\right)V_{QGP}T^{3} },
\end{multline}
with,
\begin{equation}
\mathcal{G}(\varphi ,\psi ,\frac{\mu }{T})=\mathcal{G}(0 ,0 ,\frac{\mu }{T})
+\mathcal{G}_{QG}(\varphi ,\psi ,\frac{\mu }{T}).
\label{Gfunc}
\end{equation}
The two functions are given in terms of ($T,V,\mu$) variables as follows:
\begin{equation}
\mathcal{G}(0,0,\frac{\mu }{T})=\mathbf{a}_{QG}+\frac{N_{f}N_{c}}{6\pi ^{2}}
( \frac{\mu^{4}}{2T^{4}}+\frac{\mu^{2}}{\pi ^{2}T^{2}} )
\label{G00}
\end{equation}%
and
\begin{multline}
\mathcal{G}_{QG}(\varphi ,\psi ,\frac{\mu }{T})=
\frac{\pi ^{2}d_{Q}}{24}%
\sum_{q=r,b,g}\left\{ -1+\left( \frac{\left( \alpha _{q}-i(\frac{\mu }{T}%
)\right) ^{2}}{\pi ^{2}}-1\right) ^{2}\right\} \vspace*{0.25cm} \\
-\frac{\pi ^{2}d_{G}}{24}\sum_{g=1}^{4}\left( \frac{\left( \alpha_{g}-\pi \right) ^{2}}{\pi ^{2}}-1\right) ^{2}.
\label{GQGfunc}
\end{multline}
The two factors $\mathbf{a}_{HG}$ and $\mathbf{a}_{QG}$ which are related to the degeneracies of the particles in the system are given by,
\begin{equation}
\left\{
\begin{array}{l}
\mathbf{a}_{QG}=\frac{\pi ^{2}}{12}(\frac{7}{10}d_{Q}+\frac{16}{15}d_{G}) \vspace*{0.25cm} \\
\mathbf{a}_{HG}=\frac{\pi ^{2}}{90}d_{\pi}
\end{array}
\label{AHGQGP}
\right.
\end{equation}
$d_{Q}=2N_{f}$ , $d_{G}=2$  and $d_{\pi}=3$ being the degeneracy factors of quarks,
gluons and pions respectively.$\ \alpha _{q}$ $\left( q=r,\,b,\,g\right) $ are the angles
determined by the eigenvalues of the color charge operators in eq. (\ref%
{Gfunc}):
\begin{equation}
\alpha _{r}=\frac{\varphi }{2}+\frac{\psi }{3},\;\alpha _{g}=-\frac{%
\varphi }{2}+\frac{\psi }{3},\;\alpha _{b}=-\frac{2\psi }{3},
\end{equation}%
and $\alpha _{g}$ $\left( g=1,...,4\right) $\ being: $\alpha _{1}=\alpha
_{r}-\alpha _{g},\;\alpha _{2}=\alpha _{g}-\alpha _{b},\;\alpha _{3}=\alpha
_{b}-\alpha _{r},\;\alpha _{4}=0.$
Thus, the partition function of the CQGP is then given by,
\begin{equation}
Z_{CQGP}\left( \mathbf{h}\right) =Z_{QGP}\left( \mathbf{q}\right)Z_{CC}\left( \mathbf{q}\right),
\label{ToTCQGP}
\end{equation}%
where
\begin{multline}
Z_{CC}\left( \mathbf{q}\right) =\frac{4}{9\pi ^{2}} \vspace*{0.25cm} \\
\times \int_{-\pi }^{+\pi }\int_{-\pi }^{+\pi }d\varphi d\psi M(\varphi
,\psi )e^{\mathbf{q\mathcal{G}_{QG}(\varphi ,\psi ,\frac{\mu }{T})V_{QGP}T^{3}} },
\label{ToTCC}
\end{multline}
is the colorless part and,
\begin{equation}
Z_{QGP}\left( \mathbf{h}\right) =\exp{\mathbf{(1-\mathbf{h}) VT^{3}\mathcal{G}_{QG}(0 ,0 ,\frac{\mu }{T})} }.
\label{ToTQGP}
\end{equation}%
is the QGP part without the colorless condition.
Finally the exact total partition function with the colorless condition is given by,%
\begin{equation}
Z_{TOT}\left(\mathbf{h}\right) =Z_{0}\left( \mathbf{h}\right)Z_{CC}\left( \mathbf{h}\right)
\label{ToTZ}
\end{equation}%
with,
\begin{equation}
Z_{0}\left( \mathbf{q}\right)=Z_{HG}(\mathbf{h})Z_{Vac}(\mathbf{h})Z_{QGP}(\mathbf{h}).
\end{equation}
 This latter is only the total partition function of the system without the colorless condition, which can be rewritten in its most familiar form obtained in earliest papers \cite{EarlyQGP}:
\begin{multline}
LnZ_{0}(T,V,\mu,\mathbf{h})=\Big[ \Big\{{a}_{QG}+\frac{N_{c}N_{f}}{6\pi ^{2}}\left( \pi ^{2}\frac{\mu ^{2}}{T^{2}}+\frac{\mu ^{4}}{2T^{4}} \right) \vspace*{0.25cm} \vspace*{0.25cm} \\
 -\frac{B}{T^{4}} \Big\} (1-\mathbf{h}) -{a}_{HG} \mathbf{h} \Big] VT^{3}  \label{ToTNCC}
\end{multline}
\subsection{Finite Size Hadronic Probability Density Function and $L_{mn}$%
-Method}
The definition of the Hadronic Probability Density Function in our model is
given by,
\begin{equation}
p(\mathbf{h})=\frac{Z(\mathbf{h})}{\int\limits_{0}^{1}Z(\mathbf{h})d%
\mathbf{h}}.  \label{hpdfDef}
\end{equation}
Since our hpdf is directly related to the partition function of the system,
it is believed that the whole information concerning the deconfinement
phase transition is self-contained in this hpdf. This hpdf should certainly
have different behavior in both sides of the phase transition and then we
should be able to locate the transition point just by analyzing some of its basic
properties.
Then we can perform the calculation of the mean value of any thermodynamic
quantity $\mathcal{Q}(T,\mu ,V)$ characterizing the system in the state $%
\mathbf{h}$ by,%
\begin{equation}
\langle \mathcal{Q}(T,\mu ,V)\rangle =\int\limits_{0}^{1}\mathcal{Q}\left(
\mathbf{h},T,\mu ,V\right) p\left( \mathbf{h}\right) d\mathbf{h}.
\label{mean}
\end{equation}%
In our previous work, as mentioned above, a new method was developed, which
has allowed us to calculate easily physical quantities describing
well the deconfinement phase transition to a CQGP in a finite volume $%
V $ \cite{Ladrem2005}. The most important result consists in the fact that
practically all thermal response functions calculated in this context can be simply
expressed as a function of only a certain double integral coefficient $%
L_{mn}$. The principal idea of these $L_{mn}$ has emerged in the beginning
when we performed the calculation of the $<\mathbf{h}(T,V)>$ and then
we consider that it will be very interesting if we chose the definition of $%
L_{mn} $ in a judicious way so that all thermodynamic quantities can, in one
way or the other, be written as function of these $L_{mn}$'s:
\begin{multline}
L_{m,n}\left( \mathbf{q}\right) =\int_{-\pi }^{+\pi }\int_{-\pi }^{+\pi
}d\varphi d\psi M(\varphi ,\psi )(\mathcal{G}(\varphi ,\psi,0 ))^{m}  \vspace*{0.25cm} \\
\times \frac{e^{\mathbf{q\ R}\left( \varphi ,\psi ;T,V\right) }}{\left( \mathbf{%
R}\left( \varphi ,\psi ;T,V\right) \right) ^{n}},  \label{Lmn}
\end{multline}
where the function $\mathbf{R}\left( \varphi ,\psi ;T,V\right)$ is given by,
\begin{equation}
\mathbf{R}\left( \varphi ,\psi ;T,V\right) =\left( \mathcal{G}(\varphi ,\psi,0 )-{a}_{HG}-\frac{B}{T^{4}}\right) VT^{3}.
\label{Rfunc}
\end{equation}
We can clearly see that these $L_{m,n}\left( \mathbf{q}\right) $ can be
considered as a state function depending on ($T$ ,$V$) and of course on state
variable $q,$ and they can be calculated numerically at each temperature $T$%
\ and volume $V$. As we will see later the mean value of any physical
quantity $\mathcal{Q}(T,\mu ,V)$ can therefore be calculated as a simple function of these $%
L_{m,n}\left( \mathbf{q}\right) $ evaluated \ in the hadronic phase: $%
L_{m,n}\left( 0\right) $ and in the CQGP phase: $L_{m,n}\left( 1\right) $.
An other important property of these $L_{mn}$ coefficients coefficients relies on the fact that any derivative within the $T$ \ variable and $V$ variable giving rise
to other $L_{m,n}\left( \mathbf{q}\right) $ coefficients, it is like
making a connection between different $L_{m,n}\left( \mathbf{q}\right) $
and mixing them in a simple recurrent relation \cite{LZH2011}.
\subsection{Reminder of Some Thermal Response Functions obtained previously}
The first quantity of interest for our study was the mean value of the
hadronic volume fraction $<\mathbf{h}(T,V)>$, which can be considered as
the order parameter for the phase transition investigated in this work. According to (\ref{hpdfDef}), $<\mathbf{h}(T,V)>$ has been expressed as  \cite{Ladrem2005,Herbadji2007,LZH2011}:
\begin{equation}
<\mathbf{h}(T,V)>=\frac{L_{02}\left( 1\right) -L_{02}\left( 0\right)
-L_{01}\left( 0\right) }{L_{01}\left( 1\right) -L_{01}\left( 0\right) },
\label{MOP}
\end{equation}
which shows the two limiting behaviors when approaching the thermodynamical limit :
\begin{equation}
\lim_{(T)\to \infty} {<\mathbf{h}(T,V)>=0},\qquad
\lim_{(T)\to \ 0} {<\mathbf{h}(T,V)>=1}.
\end{equation}
The asymptotic behaviors of $<\mathbf{h}(T,V)>$, can be related analytically to the Heaviside step function in the thermodynamical limit :
\begin{equation}
\lim_{(V)\to \infty} \langle \mathbf{h}\left( T,V\right)
\rangle \equiv 1-\Theta (T-T_{0}(\infty )).
\label{AsymThLim}
\end{equation}
The second quantity of interest was the energy density $\epsilon (T,V)$,
whose mean value was also calculated in the same way, and was found to be
related to $<\mathbf{h}(T,V)>$\ by the expression,%
\begin{equation}
<\epsilon (T,V)>=\frac{T^{2}}{V}<\left( \frac{\partial LnZ}{\partial T}
\right) >\label{Edensity}
\end{equation}
From our FSS analysis of the whole results, the 4FSE have been observed \cite{Ladrem2005,Herbadji2007}. These same effects have also been noticed in the present work.
We also wish to recall the definitions of the specific heat $c_{T}\left( T,V\right) =\frac{\partial
\langle \mathbf{\epsilon }\left( T,V\right) \rangle }{\partial T}$and the
thermal susceptibility $\chi _{T}\left( T,V\right) =\frac{\partial \langle
\mathbf{h}\left( T,V\right) \rangle }{\partial T}$ representing the
thermal derivatives of both $<\epsilon (T,V)>$ and $<\mathbf{h}(T,V)>$. These TRF
are very sensitive to the phase transition.
\section{Finite Size Cumulant Expansion: Theoretical Calculations}
\subsection{Definitions of the Moments, Central Moments,and Cumulants}
Let us briefly recall the standard cumulant expansion and review some of its
main properties.In probability theory and statistics, the cumulants $%
\mathbf{C}_{n}$ of a probability distribution are a set of quantities that
provide an alternative to the moments of the distribution. The moments
determine the cumulants in the sense that, any two probability distributions
whose moments are identical, have identical cumulants.
Similarly the cumulants determine the moments. In some cases theoretical
treatments of problems in terms of cumulants are simpler than those of
moments \cite{Kubo1962,ARCT2010}.
The $n^{th}$ moment of a probability density function $\ f(x)$ of a variable
$x$ is the mean value of $x^{n}$ and is mathematically defined by,
\begin{equation}
\mathbf{a}_{n}=\left\langle x^{n}\right\rangle =\int\limits_{-\infty
}^{+\infty }x^{n}f(x)dx.  \label{moment}
\end{equation}
As well known, the set of moments fully characterizes a probability density
function provided that they are all finite. At the same time the set of cumulants
that is another alternative and, for some problems is a more convenient
description. Once the set of moments are known, the probability distribution
may be obtained via reverse Fourier transform, that is the function $\Omega (t)$ which
is nothing that the mean value of the $e^{itx}$, depending only on the $t$
variable and called the \textit{characteristic function} of the distribution
$f(x)$:
\begin{equation}
\Omega(t)=\left\langle e^{itx}\right\rangle =\int\limits_{-\infty
}^{+\infty}e^{itx}f(x)d(x)= 1+\sum\limits_{n=1}^{\infty}\frac{\mathbf{a}_{n}}{n!}\left( it\right)^{n}.  \label{MomDef}
\end{equation}
So, once $\Omega (t)$ is known, all moments are known. New coefficients $%
\mathbf{C}_{n}$, which were introduced by Thiele \cite{Cram1962,THIELE}, can be
defined from the Maclaurin development of the $\ln \Omega (t)$ :
\begin{equation}
\ln \Omega (t)=\sum\limits_{n=1}^{\infty }\frac{\mathbf{C}_{n}}{n!}\left(
it\right) ^{n}.  \label{CumDef}
\end{equation}
They are called the \textit{semi-invariants} or \textit{cumulants} of the
distribution $f(x)$.
In another way when we define the central moments $\mathcal{M}_{n}$,
relatively to the mean value of $x$ ($\mathbf{a}_{1}=\left\langle
x\right\rangle $ )as,
\begin{equation}
\mathcal{M}_{n}=\int\limits_{-\infty }^{+\infty }\left( x-\mathbf{a}%
_{1}\right) ^{n}f(x)dx=\sum_{k=0}^{n}\frac{(-1)^{k}n!}{k!\left( n-k\right) !}%
\left( \mathbf{a}_{1}\right) ^{k}\mathbf{a}_{n-k}.  \label{CMvsM}
\end{equation}
Using (\ref{MomDef}), (\ref{CumDef}) and (\ref{CMvsM}) one can easily express
the cumulants $\mathbf{C}_{n}$ and the central moments $\mathcal{M}_{n}$ via the moments $\mathbf{a}_{n}$,
\vspace*{0.5cm}  % with the correct table height
\begin{equation}
\left\{
\begin{array}{l}
\mathbf{C}_{1}=\mathbf{a}_{1} \\
\mathbf{C}_{2}=\mathbf{a}_{2}-\left( \mathbf{a}_{1}\right) ^{2} \\
\mathbf{C}_{3}=\mathbf{a}_{3}-3\mathbf{a}_{1}\mathbf{a}_{2}+2\left(
\mathbf{a}_{1}\right) ^{3} \\
\mathbf{C}_{4}=\mathbf{a}_{4}-3\left( \mathbf{a}_{2}\right) ^{2}-4%
\mathbf{a}_{1}\mathbf{a}_{3}+12\left( \mathbf{a}_{1}\right) ^{2}%
\mathbf{a}_{2}-6\left( \mathbf{a}_{1}\right) ^{4} \\
\mathbf{C}_{5}=\mathbf{a}_{5}-5\mathbf{a}_{1}\mathbf{a}_{4}-10%
\mathbf{a}_{2}\mathbf{a}_{3}+20\mathbf{a}_{3}\left( \mathbf{a}%
_{1}\right) ^{2}\\
\hspace*{0.9cm}
+30\left( \mathbf{a}_{2}\right) ^{2}\mathbf{a}%
_{1}-60\left( \mathbf{a}_{1}\right) ^{3}\mathbf{a}_{2}+24(\mathbf{a}%
_{1})^{5} \\
\mathbf{C}_{6}=\mathbf{a}_{6}-6\mathbf{a}_{1}\mathbf{a}_{5}-15%
\mathbf{a}_{2}\mathbf{a}_{4}+30\mathbf{a}_{4}\left( \mathbf{a}%
_{1}\right) ^{2}-10\left( \mathbf{a}_{3}\right) ^{2}\\
\hspace*{0.85cm}
+120\mathbf{a}_{1}\mathbf{a}_{2}\mathbf{a}_{3}-120(\mathbf{a}_{1})^{3}\mathbf{a}%
_{3}+30(\mathbf{a}_{2})^{3}\\
\hspace*{0.85cm}
-270(\mathbf{a}_{1})^{2}(\mathbf{a}_{2})^{2}+360(\mathbf{a}_{1})^{4}\mathbf{a}_{2}-120(\mathbf{a}_{1})^{6}
\\
\mathbf{C}_{7}=\mathbf{a}_{7}-7\mathbf{a}_{1}\mathbf{a}_{6}-21%
\mathbf{a}_{2}\mathbf{a}_{5}+42\mathbf{a}_{1}^{2}\mathbf{a}_{5}-35%
\mathbf{a}_{3}\mathbf{a}_{4}\\
\hspace*{0.85cm}
+210\mathbf{a}_{4}\mathbf{a}_{2}%
\mathbf{a}_{1}-210\mathbf{a}_{4}\mathbf{a}_{1}^{3}+140\mathbf{a}_{1}%
\mathbf{a}_{3}^{2}+210\mathbf{a}_{3}\mathbf{a}_{2}^{2}\\
\hspace*{0.85cm}
-1260\mathbf{a}%
_{1}^{2}\mathbf{a}_{2}\mathbf{a}_{3}+840\mathbf{a}_{1}^{4}\mathbf{a}_{3}-630\mathbf{a}_{2}^{3}\mathbf{a}_{1}
+2520\mathbf{a}_{1}^{3}\mathbf{a}_{2}^{2}\\
\hspace*{0.85cm}
-2520\mathbf{a}_{1}^{5}%
\mathbf{a}_{2}+720\mathbf{a}_{1}^{7}\\
\mathbf{C}_{8}=\mathbf{a}_{8}-8\mathbf{a}_{1}\mathbf{a}_{7}-28%
\mathbf{a}_{2}\mathbf{a}_{6}+56\mathbf{a}_{1}^{2}\mathbf{a}_{6}-56%
\mathbf{a}_{3}\mathbf{a}_{5}\\
\hspace*{0.85cm}
+336\mathbf{a}_{5}\mathbf{a}_{2}%
\mathbf{a}_{1}-336\mathbf{a}_{5}\mathbf{a}_{1}^{3}-35\mathbf{a}%
_{4}^{2}+560\mathbf{a}_{1}\mathbf{a}_{4}\mathbf{a}_{3}\\
\hspace*{0.85cm}
+420\mathbf{a}_{4}%
\mathbf{a}_{2}^{2}-2520\mathbf{a}_{4}\mathbf{a}_{2}\mathbf{a}%
_{1}^{2}-1680\mathbf{a}_{1}^{4}\mathbf{a}_{4}+560\mathbf{a}_{3}^{2}%
\mathbf{a}_{2}\\
\hspace*{0.85cm}
-1680\mathbf{a}_{1}^{2}\mathbf{a}_{3}^{2}-5040\mathbf{a%
}_{3}\mathbf{a}_{2}^{2}\mathbf{a}_{1}+13440\mathbf{a}_{3}\mathbf{a}%
_{1}^{3}\mathbf{a}_{2}\\
\hspace*{0.85cm}
-6720\mathbf{a}_{1}^{5}\mathbf{a}_{3}-630\mathbf{a}_{2}^{4}+10080\mathbf{a}_{1}^{2}\mathbf{a}_{2}^{3}-25200\mathbf{a}_{1}^{4}%
\mathbf{a}_{2}^{2}\\
\hspace*{0.85cm}
+20160\mathbf{a}_{1}^{6}\mathbf{a}_{2}-5040\mathbf{%
a}_{1}^{8}\\
...........%
\end{array}%
\right. , \label{cumulant}
\end{equation}
\begin{equation}
\left\{
\begin{array}{l}
\mathcal{M}_{1}=\mathbf{0} \\
\mathcal{M}_{2}=\mathbf{a}_{2}-\left( \mathbf{a}_{1}\right) ^{2}=\sigma
^{2} \\
\mathcal{M}_{3}=\mathbf{a}_{3}-3\mathbf{a}_{1}\mathbf{a}_{2}+2\left(
\mathbf{a}_{1}\right) ^{3} \\
\mathcal{M}_{4}=\mathbf{a}_{4}-4\mathbf{a}_{1}\mathbf{a}_{3}+6\left(
\mathbf{a}_{1}\right) ^{2}\mathbf{a}_{2}-3\left( \mathbf{a}_{1}\right)
^{4} \\
\mathcal{M}_{5}=\mathbf{a}_{5}-5\mathbf{a}_{4}\mathbf{a}_{1}-10(%
\mathbf{a}_{1})^{3}\mathbf{a}_{2}+10\mathbf{a}_{3}\left( \mathbf{a}%
_{1}\right) ^{2}+4\left( \mathbf{a}_{1}\right) ^{5} \\
\mathcal{M}_{6}=\mathbf{a}_{6}-6\mathbf{a}_{5}\mathbf{a}_{1}+15(%
\mathbf{a}_{1})^{2}\mathbf{a}_{4}-20\mathbf{a}_{3}\left( \mathbf{a}%
_{1}\right) ^{3} \\
\hspace*{0.9cm}
+15\left( \mathbf{a}_{1}\right) ^{4}\mathbf{a}%
_{2}-5\left( \mathbf{a}_{1}\right) ^{6} \\
\mathcal{M}_{7}=\mathbf{a}_{7}-7\mathbf{a}_{6}\mathbf{a}_{1}+21%
\mathbf{a}_{1}{}^{2}\mathbf{a}_{5}-35\mathbf{a}_{4}\mathbf{a}%
_{1}^{3}+35\mathbf{a}_{1}^{4}\mathbf{a}_{3}\\
\hspace*{0.9cm}
-21\mathbf{a}_{1}^{5}%
\mathbf{a}_{2}-6\mathbf{a}_{1}^{7} \\
\mathcal{M}_{8}=\mathbf{a}_{8}-8\mathbf{a}_{7}\mathbf{a}_{1}+28%
\mathbf{a}_{1}{}^{2}\mathbf{a}_{6}-56\mathbf{a}_{5}\mathbf{a}%
_{1}^{3}+70\mathbf{a}_{1}^{4}\mathbf{a}_{4}\\
\hspace*{0.9cm}
-56\mathbf{a}_{1}^{5}%
\mathbf{a}_{3}+28\mathbf{a}_{1}^{6}\mathbf{a}_{2}-7\mathbf{a}_{1}^{8}\\
...........%
\end{array}%
\right. .
\end{equation}
We can also write the cumulants in terms of central moments:
\begin{equation}
\left\{
\begin{array}{l}
\mathbf{C}_{1}=\mathbf{a}_{1} \\
\mathbf{C}_{2}=\mathcal{M}_{2} \\
\mathbf{C}_{3}=\mathcal{M}_{3} \\
\mathbf{C}_{4}=\mathcal{M}_{4}-3\left( \mathcal{M}_{2}\right) ^{2} \\
\mathbf{C}_{5}=\mathcal{M}_{5}-10\mathcal{M}_{2}\mathcal{M}_{3} \\
\mathbf{C}_{6}=\mathcal{M}_{6}-15\mathcal{M}_{4}\mathcal{M}_{2}-10\mathcal{%
M}_{3}^{2}\mathcal{+}30\mathcal{M}_{2}^{3} \\
\mathbf{C}_{7}=\mathcal{M}_{7}-21\mathcal{M}_{5}\mathcal{M}_{2}-35\mathcal{%
M}_{4}\mathcal{M}_{3}\mathcal{\ +}210\mathcal{M}_{3}\mathcal{\ M}_{2}^{2} \\
\mathbf{C}_{8}=\mathcal{M}_{8}-28\mathcal{M}_{6}\mathcal{M}_{2}-56\mathcal{%
M}_{5}\mathcal{M}_{3}\mathcal{\ -}35\mathcal{\ M}_{4}^{2}\\
\hspace*{0.9cm}
\mathcal{+}420%
\mathcal{M}_{4}\mathcal{\ M}_{2}^{2}+560\mathcal{M}_{2}\mathcal{\ M}%
_{3}^{2}-630\mathcal{M}_{2}^{4} \\
...........%
\end{array}%
\right. ,
\end{equation}
which can be combined into a single recursive relationship :
\begin{equation}
\mathbf{C}_{n}=\mathcal{M}_{n}-\sum\limits_{m=1}^{n-1}C_{m-1}^{n-1}%
\mathbf{C}_{m}\mathcal{M}_{n-m}.
\end{equation}
General expressions for the connection between cumulants and moments may be
found in \cite{Risken1989}. A very convenient way to write the central moments and the
cumulants in terms of determinants,
\begin{equation}
\mathbf{C}_{n}=(-1)^{n-1}\left\vert
\begin{array}{cccccc}
\mathcal{M}_{1} & 1 & 0 & 0 & 0 & ... \\
\mathcal{M}_{2} & \mathcal{M}_{1} & 1 & 0 & 0 & ... \\
\mathcal{M}_{3} & \mathcal{M}_{2} & \binom{2}{1}\mathcal{M}_{1} & 1 & 0 & ...
\\
\mathcal{M}_{4} & \mathcal{M}_{3} & \binom{3}{1}\mathcal{M}_{2} & \binom{3}{2}%
\mathcal{M}_{1} & 1 & ... \\
\mathcal{M}_{5} & \mathcal{M}_{4} & \binom{4}{1}\mathcal{M}_{3} & \binom{4}{2}%
\mathcal{M}_{2} & \binom{4}{3}\mathcal{M}_{1} & ... \\
.... & ... & ... & ... & ... & ...%
\end{array}%
\right\vert_{n}
\end{equation}
and
\begin{equation}
\mathcal{M}_{n}=\left\vert
\begin{array}{cccccc}
\mathbf{C}_{1} & -1 & 0 & 0 & 0 & ... \\
\mathbf{C}_{2} & \mathbf{C}_{1} & -1 & 0 & 0 & ... \\
\mathbf{C}_{3} & \binom{2}{1}\mathbf{C}_{2} & \mathbf{C}_{1} & -1 & 0 &
... \\
\mathbf{C}_{4} & \binom{3}{1}\mathbf{C}_{3} & \binom{3}{2}\mathbf{C}_{2} &
\mathbf{C}_{1} & -1 & ... \\
\mathbf{C}_{5} & \binom{4}{1}\mathbf{C}_{4} & \binom{4}{2}\mathbf{C}_{3} &
\binom{4}{3}\mathbf{C}_{2} & \mathbf{C}_{1} & ... \\
.... & ... & ... & ... & ... & ...%
\end{array}%
\right\vert_{n}, \label{CMvsCum}
\end{equation}
where the determinants contain $n$ rows and $n$ columns and where $\binom{n}{k}=\frac{n!}{k!(n-k)!}$ are the standard binomial coefficients.
Then we can say that some important features of the system's partition function
can be deduced only by knowing all the moments. Each p\textnormal{\textit{th}}-order cumulant can be represented graphically as a connected cluster of p-points.
If we write the moments in terms of cumulants by inverting the relationship (\ref{cumulant}) or by expanding the determinant (\ref{CMvsCum}), the p\textnormal{\textit{th}}-order moment is then obtained by summing all possible ways to distribute the p-points into small clusters(connected or disconnected). The contribution of each way to the sum is given by the product of the connected cumulants that it represents. Due to the very important mathematical properties of the connected cumulants, it is often more convenient to work in terms of them. Henceforth and solely for simplicity, the word cumulant, implicitly means connected cumulant.
\subsection{Connected Cumulant Ratios Formalism}
In a symmetric distribution, every moment of odd order about the mean (if
exists) is evidently equal to zero. Any similar moment which is not zero may
thus be considered as a measure of the distribution's asymmetry or skewness.
The simplest of these measures is $\mathcal{M}_{3}$, which is
of the third dimension in units of the variable. In order to reduce this to
zero dimension, and so construct an absolute measure, we divide by $\mathcal{%
\sigma }^{3}.$
Reducing the fourth moment to zero dimension in the same way as above we
define the coefficient of excess (kurtosis) \ which is a measure of the
flattening\ degree of the distribution.
In the literature, other expressions of skewness and kurtosis are used
instead of what we have defined. Many other measures of skewness and
kurtosis have been proposed (see for example Pearson in \cite{Cram1962} ).
\subsubsection{General Definitions}
The Cumulants are considered as important quantities in physics but cumulant ratios are more important. Suggesting to review their definitions and deduce the most useful. Up to the present state of things, no formalism that would give the definitions of cumulant ratios in an unified way, exists. For this reason, it appeared to us instructive to try to standardize and unify the definition of the cumulant ratios in a clear and consistent way \cite{Mhamed2014}. We start by  the following definition :
\begin{equation}
\mathcal{K}_{\{\binom{j}{\beta_{j}\neq0}\}}^{\{\binom{i}{\alpha_{i}\neq0}\}} ={\prod_{j=1} \mathbf{C}_{j}^{\beta_{j}}}{\prod_{i=1} \mathbf{C}_{i}^{-\alpha_{i}}}. \\
\qquad
\label{GenDefRatioCum}
\end{equation}
which represents the generalized connected cumulant ratio between the cumulants $ \{\mathbf{C}_{j}\}$ and the cumulants $\{\mathbf{C}_{i}\}$ with positive exponents $ \{\forall i, \alpha_{i} \ge0 \text{ and } \beta_{j} \ge0 \}$ . From this definition we can distinguish four cases, namely :
\paragraph{The Normalized Cumulant Ratios} are obtained from (\ref{GenDefRatioCum}) when the following condition is fulfilled,
\begin{equation}
\sum_{i=1} \alpha_{i}\times (i)=\sum_{j=1} \beta_{j}\times (j).
\label{NormCond}
\end{equation}
\paragraph{The Unnormalized Cumulant Ratios} are those ratios in which we have the contrary case,
\begin{equation}
\sum_{i=1} \alpha_{i}\times (i)\neq \sum_{j=1} \beta_{j}\times (j).
\end{equation}
In this case we can distinguish two types of unnormalized cumulants: over-unnormalized cumulants in the case of $\displaystyle \sum_{i=1} \alpha_{i}\times (i) < \sum_{j=1} \beta_{j}\times (j)$ and under-unnormalized cumulants in the case of $\displaystyle \sum_{i=1} \alpha_{i}\times (i) > \sum_{j=1} \beta_{j}\times (j)$.
\paragraph{The p\textnormal{\textit{th}}-Order Normalized Cumulant Ratios} correspond to those in which only a p\textnormal{\textit{th}}-Order cumulant  is suitably normalized :
\begin{equation}
\forall j\in [1,m] \hspace*{0.25cm} /  \hspace*{0.25cm}  \hspace*{0.25cm} \beta_{j\neq p}=0  \texttt{ and } \beta_{p}=1
\end{equation}
thus
\begin{equation}
\mathcal{K}_{p}^{\{\binom{i}{\alpha_{i}\neq0}\}}= {\mathbf{C}_{p}} {\prod_{i=1} \mathbf{C}_{i}^{-\alpha_{i}}}. \\
\qquad
\label{GenDefCombCum}
\end{equation}
with
\begin{equation}
\sum_{i=1}^{n} \alpha_{i}\times (i)= p.
\end{equation}
\paragraph{The p\textnormal{\textit{th}}-Order Under-Normalized Cumulant Ratios} which are the most useful ones. This time, we have a particular form of the latter case, in which the indices $\{i\}$ are all less or equal to $p$ :
\begin{equation}
\mathcal{K}_{\leq p}^{\{\binom{i}{\alpha_{i}\neq0}\}}={\mathbf{C}_{p}}{\prod_{i=1}^{p} \mathbf{C}_{i}^{-\alpha_{i}}}. \\
\qquad
\label{GDefUNCum}
\end{equation}
with
\begin{equation}
\sum_{i=1}^{p} \alpha_{i}\times (i)= p,
\label{PuissCum}
\end{equation}
The numbers $\alpha_{i}$ are either integers or rational numbers. If we solve the last algebraic equation (\ref{PuissCum}), we obtain the values of ${\{\alpha_{i}\}}$ for every definition.
For example $n=4$ :
\begin{equation}
\sum_{i=1}^{4} \alpha_{i}\times (i)=\alpha_{1}+2\alpha_{2}+3\alpha_{3}+4\alpha_{4}=4.
\end{equation}
When solving this equation in the set of natural numbers $\mathbb{N}$ we find only five possibilities:
{\setlength\arraycolsep{1pt}
\begin{eqnarray}
\lbrace(\alpha_{1},\alpha_{2},\alpha_{3},\alpha_{4})=(0,2,0,0);(1,0,1,0);(0,0,0,1);{}
\nonumber\\
{}(2,1,0,0);(4,0,0,0)\rbrace{}.
\end{eqnarray}}
From the relations (\ref{GenDefRatioCum}) and (\ref{GDefUNCum}) we derive the relationship which combines two different definitions of $\textnormal{p\textit{th}}$-order under-normalized cumulant  $\mathcal{K}_{\leq p}^{\{\binom{i}{\alpha_{i}\neq0}\}}$ and \newline $\mathcal{K}_{\leq p}^{\{\binom{j}{\beta_{j}\neq0}\}}$, which is given by,
\begin{equation}
\mathcal{K}_{\leq p}^{\{\binom{i}{\alpha_{i}\neq0}\}}= \\
\mathcal{K}_{\leq p}^{\{\binom{j}{\beta_{j}\neq0}\}}\mathcal{K}_{\{\binom{j\leq p}{\beta_{j}\neq0}\}}^{\{\binom{i\leq p}{\alpha_{i}\neq0}\}}
\qquad
\label{GenRelRatioCum}
\end{equation}
with
\begin{equation}
\sum_{i=1}^{p} \alpha_{i}\times (i)= \sum_{j=1}^{p} \beta_{j}\times (j)=p.
\end{equation}
From the relation (\ref{GDefUNCum}) we see that the number of possible definitions of $\mathcal{K}_{\leq p}^{\{\binom{i}{\alpha_{i}\neq0}\}}$ increases with the order ${p}$. However, we shall not consider all definitions, but we focus only on those mostly used. Generically, the structures of all cumulants are related to each other and the behavior including the magnitudes can be deduced from the preceding.
\begin{table}
\caption{Some p-Order Under-Normalized Cumulants}
\label{tab:1}       % Give a unique label
\def\arraystretch{2}%
\begin{tabular}{|l|l|l|}
\hline
\textbf{p\textnormal{\textit{th}}-Order} & \textbf{${\{\alpha_{i}\}}$}&$\mathcal{K}_{\leq p}^{\{\binom{i}{\alpha_{i}\neq0}\}}$ \textbf {}%
\\ \hline
$2$ & ${\{\alpha_{1}=2\}}$
& $\mathbf{\mathcal{K}}_{\leq2}^{\{\binom{1}{\alpha_{1}=2}\}}=\widetilde{\sigma} ^{2}$  \\  \hline
$3$ & ${\{\alpha_{2}=3/2\}}$
& $\mathbf{\mathcal{K}}_{\leq3}^{\{\binom{2}{\alpha_{2}=3/2}\}}=\Sigma $ \\ \hline
$4$ & ${\{\alpha_{2}=2\}}$
& $\mathbf{\mathcal{K}}_{\leq4}^{\{\binom{2}{\alpha_{2}=2}\}}=\kappa $ \\ \hline
$5$ & ${\{\alpha_{2}=5/2\}}$
& $\mathbf{\mathcal{K}}_{\leq5}^{\{\binom{2}{\alpha_{2}=5/2}\}}=\Pi_{-} $ \\ \hline
$$ & ${\{\alpha_{2}=1,\alpha_{3}=1\}}$
& $\mathbf{\mathcal{K}}_{\leq5}^{\{\binom{2}{\alpha_{2}=1},\binom{3}{\alpha_{3}=1}\}}=\Pi_{+} $ \\ \hline
$6$ & ${\{\alpha_{2}=3\}}$
& $\mathbf{\mathcal{K}}_{\leq6}^{\{\binom{2}{\alpha_{2}=3}\}}=\mathcal{H}_1 $ \\ \hline
$$ & ${\{\alpha_{3}=2\}}$
& $\mathbf{\mathcal{K}}_{\leq6}^{\{\binom{3}{\alpha_{3}=2}\}}=\mathcal{H}_2 $ \\ \hline
$$ & ${\{\alpha_{2}=1,\alpha_{4}=1\}}$
& $\mathbf{\mathcal{K}}_{\leq6}^{\{\binom{2}{\alpha_{2}=1},\binom{4}{\alpha_{4}=1}\}}=\mathcal{H}_3 $ \\ \hline
$7$ & ${\{\alpha_{2}=7/2\}}$
& $\mathbf{\mathcal{K}}_{\leq7}^{\{\binom{2}{\alpha_{2}=7/2}\}}=\eta_1 $ \\ \hline
$$ & ${\{\alpha_{2}=1,\alpha_{5}=1\}}$
& $\mathbf{\mathcal{K}}_{\leq7}^{\{\binom{2}{\alpha_{2}=1},\binom{5}{\alpha_{5}=1}\}}=\eta_2 $ \\ \hline
$$ & ${\{\alpha_{2}=2,\alpha_{3}=1\}}$
& $\mathbf{\mathcal{K}}_{\leq7}^{\{\binom{2}{\alpha_{2}=2},\binom{3}{\alpha_{3}=1}\}}=\eta_3 $ \\ \hline
$$ & ${\{\alpha_{3}=1,\alpha_{4}=1\}}$
& $\mathbf{\mathcal{K}}_{\leq7}^{\{\binom{3}{\alpha_{3}=1},\binom{4}{\alpha_{4}=1}\}}=\eta_4 $ \\ \hline
$8$ & ${\{\alpha_{2}=4\}}$
& $\mathbf{\mathcal{K}}_{\leq8}^{\{\binom{2}{\alpha_{2}=4}\}}=\omega_1 $ \\ \hline
$$ & ${\{\alpha_{2}=1,\alpha_{6}=1\}}$
& $\mathbf{\mathcal{K}}_{\leq8}^{\{\binom{2}{\alpha_{2}=1},\binom{6}{\alpha_{6}=1}\}}=\omega_2 $ \\ \hline
$$ & ${\{\alpha_{2}=2,\alpha_{4}=1\}}$
& $\mathbf{\mathcal{K}}_{\leq8}^{\{\binom{2}{\alpha_{2}=2},\binom{4}{\alpha_{4}=1}\}}=\omega_3 $ \\ \hline
$$ & ${\{\alpha_{3}=1,\alpha_{5}=1\}}$
& $\mathbf{\mathcal{K}}_{\leq8}^{\{\binom{3}{\alpha_{3}=1},\binom{5}{\alpha_{5}=`}\}}=\omega_4 $ \\ \hline
$$ & ${\{\alpha_{2}=1,\alpha_{3}=2\}}$
& $\mathbf{\mathcal{K}}_{\leq8}^{\{\binom{2}{\alpha_{2}=1},\binom{3}{\alpha_{3}=2}\}}=\omega_5 $ \\ \hline
$$ & ${\{\alpha_{4}=2\}}$
& $\mathbf{\mathcal{K}}_{\leq8}^{\{\binom{4}{\alpha_{4}=2}\}}=\omega_6 $ \\ \hline
\end{tabular}%
\vspace*{0.2cm}  % with the correct table height
\end{table}
\subsubsection{The First Order Under-Normalized Cumulant Ratio: Normalized Mean Value}
Because the first cumulant is the mean value of $x$ :$\mathbf{C}_{1}=\mathbf{a}%
_{1}=\left\langle x\right\rangle $ then the first order under-normalized cumulant ratio is $\mathcal{K}_{\leq 1}^{\{\binom{1}{\alpha_{1}=1}\}}=1$
\subsubsection{The Second Order Under-Normalized Cumulant Ratio : Normalized Variance}
The second order under-normalized cumulant ratio may be referred to the \textit{normalized variance} and defined as,
\begin{equation}
\mathbf{\mathcal{K}_{\leq 2}^{\{\binom{1}{\alpha_{1}=2}\}}}=\frac{\sigma ^{2}}{\langle x\rangle ^{2}}=\frac{\mathbf{C}_{2}}{\mathbf{C}_{1}^{2}}=\frac{\mathcal{M}_{2}}{\mathbf{C}_{1}^{2}}=
\frac{\langle x^{2}\rangle}{\langle x\rangle ^{2}}-1 .
\label{variance1}
\end{equation}
\subsubsection{The Third Order Under-Normalized Cumulant Ratio: Skewness}
The third order under-normalized cumulant ratio is a measure of symmetry, or more precisely, the lack of
symmetry. A distribution, or data set, is symmetric if it looks the same to
the left and the right of the center point. The $3^{rd}$\ cumulant for a normal
distribution is zero, and any symmetric distribution will have a third
central moment, if defined, near zero. Then the $3^{rd}$\ under-normalized cumulant ratio is called the \textit{Skewness} $\Sigma $ and is defined as,
\begin{equation}
\mathbf{\mathcal{K}}_{\leq3}^{\{\binom{2}{\alpha_{2}=3/2}\}}=\Sigma =\frac{\mathbf{C}_{3}}{\left( \mathbf{C}_{2}\right) ^{3/2}}=\frac{%
\mathcal{M}_{3}}{\mathcal{M}_{2}^{3/2}}.  \label{skewness}
\end{equation}
A distribution is skewed to the left (the tail of the distribution is
heavier on the left) will have a negative skewness. A distribution that is
skewed to the right (the tail of the distribution is heavier on the right)
will have a positive skewness.
\subsubsection{The Fourth Order Under-Normalized Cumulant Ratio: Kurtosis}
The fourth order under-normalized cumulant ratio is a measure of whether the distribution is peaked
or flat relatively to a normal distribution. Since it is the expectation value to the fourth power, the fourth central moment, where defined, is always positive.
Because the fourth cumulant of a normal distribution is $3\sigma ^{4}$, then the most commonly definition of the fourth order under-normalized cumulant ratio called \textit{Kurtosis } $\mathcal{\kappa}$ is
\begin{equation}
\mathbf{\mathcal{K}}_{\leq4}^{\{\binom{2}{\alpha_{2}=2}\}}=\kappa ={\frac{\mathbf{C}_{4}}{\left(
\mathbf{C}_{2}\right) ^{2}}=}\frac{\mathcal{M}_{4}}{\mathcal{M}_{2}^{2}}-3,
\label{kurtosis}
\end{equation}
so that the standard normal distribution has a kurtosis of zero. Positive
kurtosis indicates a "peaked" distribution and negative kurtosis indicates a
"flat" distribution. Following the classical interpretation, kurtosis
measures both the "peakedness" of the distribution and the heaviness of its
tail \cite{Kurtosis1988}.
In addition to this, Binder was the first who has proposed and studied the fourth cumulant as it was defined in \cite{CLB1986}using the moments of the energy probability distribution:
\begin{equation}
\mathcal{B}_{4}=1-\frac{\mathbf{a}_{4}}{3\ \mathbf{a}_{2}^{2}}.  \label{binder4}
\end{equation}
This was introduced as a quantity whose behavior could determine the order of the phase transition.
If we replace the moments by the central moments, we get another completely different physical quantity, which is related the kurtosis as,
\begin{equation}
\mathcal{B}^{c}_{4}=1-\frac{\mathcal{M}_{4}}{3\ \mathcal{M}_{2}^{2}}=-\frac{1}{%
3\ }\,\mathcal{\kappa }.  \label{cbinder4}
\end{equation}
and can be easily derived from our general definition of connected cumulant ratios (\ref{GenDefRatioCum}). This new cumulant, as we have mentioned before, is called connected Binder cumulant or conventional Binder cumulant. However, to avoid confusion in the appellations we simply keep the name of Binder cumulant for the first quantity. Historically, this new cumulant was first introduced and studied by Binder in 1984 \cite{BL1984}. Seven years later, this new cumulant was reconsidered in an independant and important paper by Lee and Kosterlitz in the context of a different model \cite{Lee1991}. The difference between the two Binder cumulants attracted little attention in its early years. But, in 1993, Janke has illuminated the most important difference in a comparative and fruitful study between the two cumulants \cite{Janke1993}. The great significance of the connected Binder cumulant relative to the Binder cumulant is summed up in the following points:(1) the thermal behaviors of two Binder cumulants are very different, particularly in the transition region,(2) the connected Binder cumulant has a richer structure than the Binder cumulant, (3) the connected Binder cumulant is more efficient in locating the true finite volume transition point than the Binder cumulant.
This cumulant is a finite size scaling function \cite{BL1984,CLB1986,Binder1997,LB2000,BH1988},
and is widely used to indicate the order of the transition in a finite volume.
In ordered systems, a good parameter to locate phase transitions is exactly this
connected Binder cumulant which is the kurtosis of the order-parameter probability
distribution. The uniqueness of the ground state in that case is enough to
guarantee that the Binder cumulant takes the universal value at zero
temperature for any finite volume.
\subsubsection{The Fifth Order Under-Normalized Cumulant Ratios: Pentosis}
The fifth order under-normalized cumulant ratio, which is called \textit{Pentosis}, can be defined in two
ways. The first one is given by,
\begin{equation}
\mathbf{\mathcal{K}}_{\leq5}^{\{\binom{2}{\alpha_{2}=1},\binom{3}{\alpha_{3}=1}\}}={\Pi }_{+}{=\frac{\mathbf{C}_{5}}{\mathbf{C}_{2}%
\mathbf{C}_{3}}=}\frac{\mathcal{M}_{5}}{\mathcal{M}_{2}\mathcal{M}_{3}}-10%
\label{pentosis1}
\end{equation}
and the second definition is given by,
\begin{equation}
\mathbf{\mathcal{K}}_{\leq5}^{\{\binom{2}{\alpha_{2}=5/2}\}}={\Pi }_{-}{=\frac{\mathbf{C}_{5}}{\left( \mathbf{C}%
_{2}\right) ^{5/2}}}=\frac{\mathcal{M}_{5}}{\mathcal{M}_{2}^{5/2}}-10\Sigma .
\label{P-}
\end{equation}
The two forms of pentosis are of course related by,
\begin{equation}
{\Pi }_{-}=\Sigma {\Pi }_{+}.
\label{P+}
\end{equation}
a relation that we can deduce from the general relationship (\ref{GenRelRatioCum}).
\subsubsection{The Sixth Order Under-Normalized Cumulant Ratios: Hexosis}
The sixth order under-normalized cumulant ratio is, analogously to
Pentosis and Kurtosis, coined Hexosis. It can be defined in one of the following ways \cite{CumBio1994,Mhamed2014}:
\begin{equation}
\mathbf{\mathcal{K}}_{\leq6}^{\{\binom{2}{\alpha_{2}=3}\}}=\mathcal{H}_{1}=\frac{\mathbf{C}_{6}}{\left( \mathbf{C}_{2}\right)
^{3}}=\frac{\mathcal{M}_{6}-15\mathcal{M}_{4}\mathcal{M}_{2}\mathcal{-}10%
\mathcal{M}_{3}^{2}}{\mathcal{M}_{2}^{3}}+30,
\label{H1}
\end{equation}
\begin{equation}
\mathbf{\mathcal{K}}_{\leq6}^{\{\binom{3}{\alpha_{3}=2}\}}=\mathcal{H}_{2}=\frac{\mathbf{C}_{6}}{\left(\mathbf{C}_{3}\right) ^{2}}=\frac{\mathcal{M}%
_{6}-15\mathcal{M}_{4}\mathcal{M}_{2}+30\mathcal{M}_{2}^{3}}{\mathcal{M}%
_{3}^{2}}-10{\qquad }  \label{H2}
\end{equation}
\begin{equation}
\mathbf{\mathcal{K}}_{\leq6}^{\{\binom{2}{\alpha_{2}=1},\binom{4}{\alpha_{4}=1}\}}=\mathcal{H}_{3}=\frac{\mathbf{C}_{6}}{\mathbf{C}_{4}\mathbf{C}_{2}}=%
\frac{\mathcal{M}_{6}-10\mathcal{M}_{3}^{2}\mathcal{-}15\mathcal{M}_{2}^{3}}{%
\left[ \mathcal{M}_{4}-3\left( \mathcal{M}_{2}\right) ^{2}\right] \mathcal{M}%
_{2}}-15{\qquad }.
\label{H3}
\end{equation}
It is easy to show that the three definitions of hexosis are related each other by the relations,
\begin{equation}
\mathcal{H}_{3}= \mathcal{\kappa }^{-1} \mathcal{H}_{1}= {\Sigma }^{2} \mathcal{\kappa}^{-1} \mathcal{H}_{2},
\label{H123}
\end{equation}
which can be deduced from the general relationship (\ref{GenRelRatioCum}).
\subsubsection{The Seventh Order Under-Normalized Cumulant Ratios: Heptosis}
With the same spirit and by analogy to pentosis, kurtosis and hexosis, we can term the seventh order under-normalized cumulant ratio as Heptosis \cite{Mhamed2014} $\eta$.
One of the possible definition of heptosis is given by,
\begin{multline}
\mathbf{\mathcal{K}}_{\leq7}^{\{\binom{2}{\alpha_{2}=2},\binom{3}{\alpha_{3}=1}\}}=\eta_3=
\frac{\mathbf{C}_{7}}{\mathbf{C}_{2}^{2}\mathbf{C}_{3}}= \vspace*{0.25cm} \\
\frac{\mathcal{M}_{7}}{\mathcal{M}_{2}^{2}\mathcal{M}_{3}}
-21\frac{\mathcal{M}_{5}}{\mathcal{M}_{2}\mathcal{M}_{3}}
-35\frac{\mathcal{M}_{4}}{\mathcal{M}_{2}^{2}}+210,
\label{heptosis}
\end{multline}
\subsubsection{The Eighth Order Under-Normalized Cumulant Ratios: Octosis}
Concerning the eighth order under-normalized cumulant ratio, which can be termed Octosis \cite{Mhamed2014} and take one of the eight definitions from table (\ref{tab:1}),
\begin{multline}
\mathbf{\mathcal{K}}_{\leq8}^{\{\binom{2}{\alpha_{2}=1},\binom{3}{\alpha_{3}=2}\}}=
\omega_5=\frac{\mathbf{C}_{8}}{\mathbf{C}_{2}\mathbf{C}_{3}^{2}}=\frac{\mathcal{M}_{8}}{\mathcal{M}_{3}^{2}\mathcal{M}_{2}}
-28\frac{\mathcal{M}_{6}}{\mathcal{M}_{3}^{2}} \vspace*{0.25cm} \\
-56\frac{\mathcal{M}_{5}}{\mathcal{M}_{3}\mathcal{M}_{2}}-35\frac{\mathcal{M}_{4}^{2}}{\mathcal{M}_{3}^{2}\mathcal{M}_{2}}
+420\frac{\mathcal{M}_{4}\mathcal{M}_{2}}{\mathcal{M}_{3}^{2}}
-630\frac{\mathcal{M}_{2}^{3}}{\mathcal{M}_{3}^{2}}+560.
\label{Octosis}
\end{multline}
\subsubsection{Three Unnormalized Cumulant Ratios}
We are also interested in studying different unnormalized combinations of the cumulants. Their importance was revealed and emphasized in several recent works \cite{KR2011,Dai2012,Chen2012,Step2009}. The first combination contains the variance $\mathcal{\sigma }^{2}$, kurtosis $\mathcal{%
\kappa }$ and skewness $\mathbf{\Sigma }$ and is defined as,
\begin{equation}
\mathcal{O} =\frac{\mathcal{\sigma }^{2}
\mathcal{\kappa } }{\mathbf{\Sigma }}=\frac{\mathbf{C}_{2}^{1/2}\mathbf{C}_{4}}{\mathbf{C}_{3}}%
=\mathcal{K}_{\{\binom{1}{3}\}}^{\{\binom{1}{4}\},\{\binom{1/2}{2}\}}=\frac{\mathcal{M}_{2}^{\frac{1}{2}}\left(
\mathcal{M}_{4}-3\mathcal{M}_{2}^{2}\right) }{\mathcal{M}_{3}}.
\label{ODef}
\end{equation}
The second one contains only the variance $\mathcal{\sigma }^{2}$ and skewness $\mathbf{\Sigma }$ and is given by,
\begin{equation}
\mathcal{U} =\frac{1}{\mathcal{\sigma }^{2} \mathbf{\Sigma }}=\frac{\mathbf{C}_{2}^{1/2}}{\mathbf{C}_{3}}%
=\mathcal{K}_{\{\binom{1}{3}\}}^{\{\binom{1/2}{2}\}}=\frac{\mathcal{M}_{2}^{\frac{1}{2}}}{\mathcal{M}_{3}}
{\qquad }.
\label{UDef}
\end{equation}
However,the third combination contains the variance $\mathcal{\sigma }^{2}$ and kurtosis $\mathcal{%
\kappa }$ and is given by,
\begin{equation}
\mathcal{N}=\mathcal{\sigma }^{2}
\mathcal{\kappa } =\frac{\mathbf{C}_{4}}{\mathbf{C}_{2}}%
=\mathcal{K}_{\{\binom{1}{2}\}}^{\{\binom{1}{4}\}}=\frac{\mathcal{M}_{4}-3\mathcal{M}_{2}^{2}}{%
\mathcal{M}_{2}}
{\qquad }.
\label{NDef}
\end{equation}
\subsection{Finite size cumulant expansion of the hadronic probability
density function $p(\mathbf{h})$ as function of $L_{mn}(q,T,V)$}
Using our hadronic probability density function $p(\mathbf{h})$, we derive
the general expression of the mean value $\left\langle \mathbf{h}^{n}\right\rangle$ as a function of $%
L_{mn}(q,T,V)$ \cite{GRY2007}. Afterwards, one can express the different cumulants $\mathbf{C}_{n}(T,V)$ in terms of these $L_{mn}(q,T,V)$ using (\ref{CMvsM}) and (\ref{CMvsCum}). Keeping in mind that these double integrals
$L_{mn}(q,T,V)$ are state functions depending on the temperature $T$, on the
volume $V$ and on the state variable $q$. One can hide their dependence on $%
(T,V)$ just to avoid overloading relationships. After some algebra, we get the result,
\begin{equation}
\left\langle \mathbf{h}^{n}\right\rangle (T,V)=\frac{n!L_{0,n+1}\left(
1\right) -\sum_{k=0}^{n} \binom{n}{k} k!L_{0,k+1}\left( 0\right) }{L_{0,1}\left(
1\right) -L_{0,1}\left( 0\right) }.  \label{meanhnLmn}
\end{equation}
Using this general expression of the mean value and from (\ref{cumulant}) we derive the six first cumulants(see appendix),
\begin{equation}
\left\{
\begin{array}{l}
\mathbf{C}_{1}(T,V)=\langle \mathbf{h}\rangle \\
\mathbf{C}_{2}(T,V)=\left\langle \mathbf{h}^{2}\right\rangle-\left\langle \mathbf{h}\right\rangle ^{2} \\
\mathbf{C}_{3}(T,V)=\left\langle \mathbf{h}^{3}\right\rangle-3\left\langle \mathbf{h}\right\rangle \left\langle \mathbf{h}^{2}\right\rangle +2\left\langle \mathbf{h}\right\rangle ^{3} \\
\mathbf{C}_{4}(T,V)=\left\langle \mathbf{h}^{4}\right\rangle-3\left\langle \mathbf{h}^{2}\right\rangle ^{2}-4%
\left\langle \mathbf{h}\right\rangle\left\langle \mathbf{h}^{3}\right\rangle+12 \left\langle \mathbf{h}\right\rangle ^{2}%
\left\langle \mathbf{h}^{2}\right\rangle \\
\hspace*{1.9cm}
-6 \left\langle \mathbf{h}\right\rangle^{4}\\
\mathbf{C}_{5}(T,V)=\left\langle \mathbf{h}^{5}\right\rangle-5\left\langle \mathbf{h}\right \rangle\left\langle \mathbf{h}^{4}\right\rangle-10%
\left\langle \mathbf{h}^{2}\right\rangle \left\langle \mathbf{h}^{3}\right\rangle\\
\hspace*{1.9cm}
+20\left\langle \mathbf{h}^{3}\right\rangle  \left\langle \mathbf{h}\right\rangle ^{2}+30 \left\langle \mathbf{h}^{2}\right\rangle ^{2}\left\langle \mathbf{h}\right\rangle \\
\hspace*{1.85cm}
-60 \left\langle \mathbf{h}\right\rangle ^{3}\left\langle \mathbf{h}^{2}\right\rangle +24\langle \mathbf{h} \rangle^{5} \\
\mathbf{C}_{6}(T,V)=\left\langle \mathbf{h}^{6}\right\rangle -6\left\langle \mathbf{h}\right\rangle \left\langle \mathbf{h}^{5}\right\rangle -15%
\left\langle \mathbf{h}^{2}\right\rangle \left\langle \mathbf{h}^{4}\right\rangle \\
\hspace*{1.85cm}
+30\left\langle \mathbf{h}^{4}\right\rangle  \left\langle \mathbf{h}\right\rangle ^{2}-10 \left\langle \mathbf{h}^{3}\right\rangle ^{2}+120\left\langle \mathbf{h}\right\rangle \left\langle \mathbf{h}^{2}\right\rangle \left\langle \mathbf{h}^{3}\right\rangle \\
\hspace*{1.85cm}
-120 \left\langle \mathbf{h}\right\rangle ^{3}\left\langle \mathbf{h}^{3}\right\rangle +30 \left\langle \mathbf{h}^{2}\right\rangle^{3}-270\langle \mathbf{h} \rangle^{2}\langle \mathbf{h}^{2}\rangle ^{2} \\
\hspace*{1.85cm}
+360 \langle \mathbf{h} \rangle ^{4}\left\langle \mathbf{h}^{2}\right\rangle-120 \langle \mathbf{h} \rangle ^{6}\\
...........%
\end{array}%
\right. , \label{hcumulant}
\end{equation}
Afterwards we derive the final expression of both p-order under-normalized and unnormalized cumulants under consideration.
The first cumulant is none other than the order parameter $\left\langle \mathbf{h}%
\right\rangle (T,V)$ and is given by (\ref{OPLmn}).
The variance $\sigma ^{2}\left( T,V\right) $ is given by,
\begin{equation}
\mathbf{\sigma }^{2}\left( T,V\right) =\left[ \langle \mathbf{h}%
^{2}\rangle -\langle \mathbf{h}\rangle ^{2}\right].  \label{VarianceLmn}
\end{equation}
The skewness $\Sigma \left( T,V\right) $, is given by,
\begin{equation}
\mathbf{\Sigma }\left( T,V\right) =\frac{\langle \left( \mathbf{h}-\langle
\mathbf{h}\rangle \right) ^{3}\rangle }{\sigma ^{3}}=\frac{\left[ \langle
\mathbf{h}^{3}\rangle -3\langle \mathbf{h}\rangle \langle \mathbf{h}%
^{2}\rangle +2\langle \mathbf{h}\rangle ^{3}\right] }{\left[ \langle
\mathbf{h}^{2}\rangle -\langle \mathbf{h}\rangle ^{2}\right] ^{3/2}},
\label{SkwenessLmn}
\end{equation}
and the kurtosis $\mathcal{\kappa }\left( T,V\right) $is given by,
\begin{multline}
\mathcal{\kappa }\left( T,V\right) =\frac{\langle \left( \mathbf{h}%
-\langle \mathbf{h}\rangle \right) ^{4}\rangle }{\sigma ^{4}}-3= \vspace*{0.25cm} \\
\frac{\left[ \langle \mathbf{h}^{4}\rangle -4\langle \mathbf{h}\rangle \langle
\mathbf{h}^{3}\rangle -6\langle \mathbf{h}\rangle ^{4}+12\langle
\mathbf{h}\rangle ^{2}\langle \mathbf{h}^{2}\rangle -3\langle \mathbf{h%
}^{2}\rangle ^{2}\right] }{\left[ \langle \mathbf{h}^{2}\rangle -\langle
\mathbf{h}\rangle ^{2}\right] ^{2}}.
\label{KurtosisLmn}
\end{multline}
And finally the pentosis ${\Pi }_{+}\left( T,V\right) $, which is
given by,
\begin{equation}
\left\{
\begin{array}{l}
\hspace{1.5cm}\mathbf{\Pi }_{+}\left( T,V\right) =\mathbf{N1}/\mathbf{D1} \\
\mathbf{N1}=
\left[
\langle \mathbf{h}^{5}\rangle -5\langle \mathbf{h}^{4}\rangle \langle
\mathbf{h}\rangle +20\langle \mathbf{h}^{3}\rangle \langle \mathbf{h}%
\rangle ^{2}-60\langle \mathbf{h}^{2}\rangle \langle \mathbf{h}\rangle^{3} \right. \\
\hspace*{2.0cm}
\left.
-10\langle \mathbf{h}^{2}\rangle \langle \mathbf{h}^{3}\rangle
+30\langle \mathbf{h}^{2}\rangle ^{2}\langle \mathbf{h}\rangle
+24\langle \mathbf{h}\rangle ^{5}\right]  \\
\mathbf{D1}=
\left[ \langle \mathbf{h}%
^{2}\rangle \langle \mathbf{h}^{3}\rangle -3\langle \mathbf{h}\rangle
\langle \mathbf{h}^{2}\rangle ^{2}+5\langle \mathbf{h}^{2}\rangle
\langle \mathbf{h}\rangle ^{3}- \right. \\
\hspace*{5.0cm}
\left. \langle \mathbf{h}\rangle ^{2}\langle
\mathbf{h}^{3}\rangle -2\langle \mathbf{h}\rangle ^{5}\right]\\
\end{array}%
\right. .
\label{hP+}
\end{equation}
And finally the hexosis $\mathcal{H}_{1}(T,V)$, which is given by,
\begin{equation}
\left\{
\begin{array}{l}
\hspace{1.5cm}\mathcal{H}_{1}\left( T,V\right) =\mathbf{N2}/\mathbf{D2} \\
\mathbf{N2}=
\left[\langle \mathbf{h}^{6}\rangle -6\langle \mathbf{h}^{5}\rangle
\langle \mathbf{h}\rangle -15\langle \mathbf{h}^{2}\rangle\langle \mathbf{h}^{4}\rangle+30\langle \mathbf{h}\rangle^{2}\langle \mathbf{h}^{4}\rangle \right. \\
\hspace*{0.6cm}
\left. -10\langle \mathbf{h}^{3}\rangle^{2}+120\left\langle \mathbf{h}\right\rangle \left\langle \mathbf{h}^{2}\right\rangle \left\langle \mathbf{h}^{3}\right\rangle-120 \left\langle \mathbf{h}\right\rangle ^{3}\left\langle \mathbf{h}^{3}\right\rangle  \right. \\
\hspace*{0.6cm}
\left. +30 \left\langle \mathbf{h}^{2}\right\rangle^{3} -270\langle \mathbf{h} \rangle^{2}\langle \mathbf{h}^{2}\rangle ^{2} +360 \langle \mathbf{h} \rangle ^{4}\left\langle \mathbf{h}^{2}\right\rangle-120 \langle \mathbf{h} \rangle ^{6} \right] \\
\mathbf{D2}=
\left[\langle \mathbf{h}^{2}\rangle^{3} +3\langle \mathbf{h}\rangle^{4}
\langle \mathbf{h}^{2}\rangle -3\langle \mathbf{h}^{2}\rangle^{2}\langle \mathbf{h}\rangle ^{2}-\langle \mathbf{h}\rangle ^{6} \right]
\end{array}%
\right. .
\label{hH1}
\end{equation}
The expressions of ${\Pi }_{-}\left( T,V\right) $ and $\mathcal{H}_{2,3}(T,V)$ can be derived easily from those of (\ref{P+}) and (\ref{H1}) using (\ref{H123}).
Let us now go to the unnormalized cumulants as defined in(\ref{ODef},\ref{UDef},\ref{NDef}). Final expressions of $\mathcal{O,U,N}$ are :
\begin{multline}
\mathcal{O}\left( T,V\right) =\frac{\mathcal{\sigma }^{2}\left( T,V\right)
\mathcal{\kappa }\left( T,V\right) }{\mathbf{\Sigma }\left( T,V\right) }= \vspace*{0.25cm} \\
\left[ \langle \mathbf{h}^{4}\rangle -4\langle \mathbf{h}\rangle
\langle \mathbf{h}^{3}\rangle -6\langle \mathbf{h}\rangle ^{4}+12\langle
\mathbf{h}\rangle ^{2}\langle \mathbf{h}^{2}\rangle -3\langle \mathbf{h%
}^{2}\rangle ^{2}\right] \vspace*{0.3cm} \\
\times \frac{\left[ \langle \mathbf{h}^{2}\rangle -\langle
\mathbf{h}\rangle ^{2}\right] ^{1/2}}{\left[ \langle \mathbf{h}%
^{3}\rangle -3\langle \mathbf{h}\rangle \langle \mathbf{h}^{2}\rangle
+2\langle \mathbf{h}\rangle ^{3}\right] }
\label{OLmn}
\end{multline}
\begin{equation}
\mathcal{U}\left( T,V\right) =\frac{1}{\mathcal{\sigma }^{2}\left(
T,V\right) \mathbf{\Sigma }\left( T,V\right) }=\frac{\left[ \langle
\mathbf{h}^{2}\rangle -\langle \mathbf{h}\rangle ^{2}\right] ^{1/2}}{%
\left[ \langle \mathbf{h}^{3}\rangle -3\langle \mathbf{h}\rangle \langle
\mathbf{h}^{2}\rangle +2\langle \mathbf{h}\rangle ^{3}\right] }
\label{ULmn}
\end{equation}
and
\begin{multline}
\mathcal{N}\left( T,V\right) = \mathcal{\sigma }^{2}\left( T,V\right)
\mathcal{\kappa }\left( T,V\right)= \vspace*{0.25cm} \\
\frac{\left[ \langle \mathbf{h}%
^{4}\rangle -4\langle \mathbf{h}\rangle \langle \mathbf{h}^{3}\rangle
-6\langle \mathbf{h}\rangle ^{4}+12\langle \mathbf{h}\rangle ^{2}\langle
\mathbf{h}^{2}\rangle -3\langle \mathbf{h}^{2}\rangle ^{2}\right] }{%
\left[ \langle \mathbf{h}^{2}\rangle -\langle \mathbf{h}\rangle ^{2}%
\right] }. \label{NLmn}
\end{multline}
We will see after studying these new thermodynamic functions,
that their FSS analysis will allow to identify the transition region,
to define judiciously the finite volume transition point and analyse its
behavior when approaching the thermodynamic limit.
\section{Finite Size Cumulant Expansion : Results and Discussion}
Firstly, one may notice a clear sensitivity, of the whole quantities studied in this work,
to finite volume of the system. Exactly as in the case of the results obtained in our previous work \cite{Ladrem2005,LZH2011}, the 4FSE cited above are observed.
\begin{figure}
\resizebox{0.55\textwidth}{!}{%
  \includegraphics{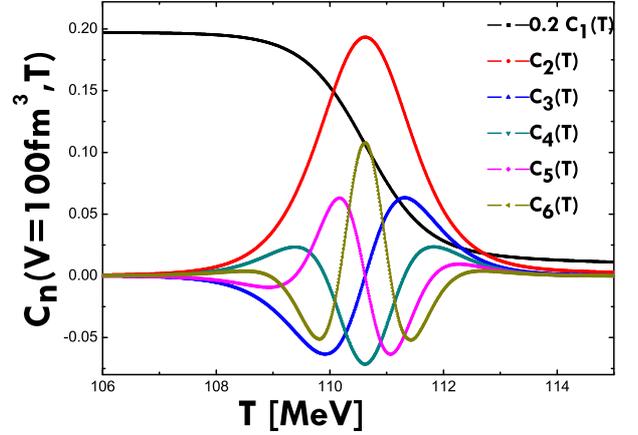}
}
\caption{Behavior of Different Cumulants $C_{n=1,2,3,4,5,6}(T,V)$ vs Temperature for Volume $=1000fm^{3}$.}
\label{diffcum}       % Give a unique label
\end{figure}
\begin{figure}
\resizebox{0.50\textwidth}{!}{%
  \includegraphics{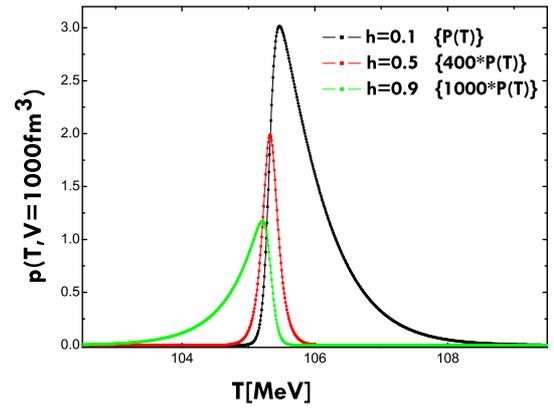}
}
\caption{Hadronic Probability Distribution Function $p(\mathbf{h},T,V=1000fm^{3})$ vs Temperature for different values of $\mathbf{h}=0.1,0.5,0.9$}
\label{hpdfplot}       % Give a unique label
\end{figure}
\begin{figure}
\resizebox{0.40\textwidth}{!}{%
  \includegraphics{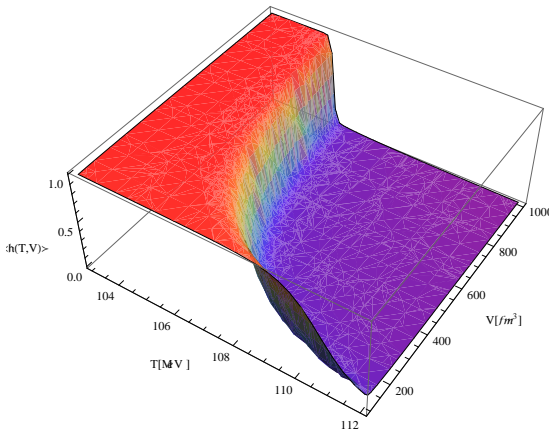}
}
\caption{3-Dim plot of the Order Parameter $<\mathbf{h}(T,V)>$ vs Temperature and Volume.}
\label{order3dim}       % Give a unique label
\end{figure}
\begin{figure}
%
%\resizebox{0.50\textwidth}{!}{%
%  \includegraphics{Figure033.eps}
%}
\caption{3-Dim plot of Variance ${\sigma }^{2}\left( T,V\right)$ vs Temperature and Volume.}
\label{fig:4}       % Give a unique label
\end{figure}
\begin{figure}
%
%\resizebox{0.50\textwidth}{!}{%
%  \includegraphics{Figure034.eps}
%}
\caption{3-Dim plot of Skewness ${\Sigma }\left( T,V\right)$ vs Temperature and Volume.}
\label{skew3dim}       % Give a unique label
\end{figure}
\begin{figure}
%
%\resizebox{0.50\textwidth}{!}{%
%  \includegraphics{Figure035.eps}
%}
\caption{3-Dim plot of Kurtosis ${\kappa }\left( T,V\right)$ vs Temperature and Volume}
\label{kurt3dim}       % Give a unique label
\end{figure}
The variation of the different cumulants and cumulant ratios versus temperature are
illustrated in Figs. (1-13) respectively, for various finite sizes. They show
interesting features. It can be clearly seen that the different finite peaks
appearing in the different quantities have width $\delta T\left( V\right) $
becoming small when approaching the thermodynamic limit. This result is
expected, since the order parameter looks like a step function when the
volume $V$ goes to infinity, as it is well known.
The rounding of the cumulants behavior is a
consequence of the finite size effects of the bulk
singularity. We notice in all curves, the emergence of a transition region, roughly bounded by two particular points, which narrows as the volume increases. In this region, all thermodynamical quantities present an oscillatory behavior which becomes faster when approaching the thermodynamic limit.
Our previous works \cite{Ladrem2005,LZH2011} have shown that both $%
\langle \mathbf{h}\rangle $ and $\frac{<\varepsilon >}{T^{4}}$ exhibit a
finite sharp discontinuity, which is related to the latent heat of the
deconfinement phase transition, at bulk transition temperature $%
T_{0}\left( \infty \right) =\left[ \frac{90B}{34\pi ^{2}}\right]
^{1/4}=104.34796MeV$, reflecting the first order character of the
phase transition. It is well known that the latent heat is the amount of energy density
necessary to convert one phase into the other at the transition point. In our
case, the latent heat can be calculated:$L_{H}\left(
\infty \right) =4B$.
This finite discontinuity can be mathematically described by a step
function, which transforms to a $\delta $-function in $\chi_{T} $ and $c_{T}$. When
the volume decreases, all quantities vary continuously such that the finite
sharp jump is rounded off and the $\delta $-peaks are smeared out into
finite peaks over a range of temperature $\delta T\left( V\right) $. Physically, we can interpret these 4FSE as due to the finite probability of presence of the CQGP phase below the transition point and of the hadron phase above it, induced by the considerable thermodynamical fluctuations.
In Fig. (1), we show the plot of the first six cumulants as functions of
temperature at fixed volume :$1000fm^{3}$. A multiple peaks structure can be observed on these curves, except in the case of the first cumulant  $C_{1}(T)$. For each additional order, a new hump (peak)is introduced. These peaks are broadened, smaller is the volume.
Also, we notice that the inflection point in the first cumulant $C_{1}(T)$ becomes a maximum point for the second order cumulant $C_{2}(T)$, a zero point in the third cumulant $C_{3}(T)$ and so on.
The number of times that a given cumulant changes its sign, is directly related to the order of the cumulant.
The sign change for the cumulants starts at the third one. It happens twice in the fourth, thrice in the fifth and four times in the sixth order cumulants. The common feature is that the higher the order of the cumulant, the higher frequency of the fluctuation pattern is.
Also, we notice that all cumulants have the same vanishing value at
low or high temperatures. In the middle region, which in principle is
considered as the transition region, the value of the cumulants
presents an oscillatory behavior due to the thermodynamical fluctuations during the phase transition.
When we carefully analyse the behavior of the hpdf for different values of
$\mathbf{h}=0.1,0.5$ and $0.9$ on Fig.2, we note that in the case of $\mathbf{h}=0.5$ the hpdf
looks like very symmetric and for these reasons we expect the skewness to be
zero. The hpdf distribution is skewed right before the transition $\mathbf{h}=0.1$
and becomes skewed left after the occurrence of the phase transition $\mathbf{h}=0.9$. The
peaks of the hpdf are more pronounced when we go from a pure CQGP phase to
a pure hadronic phase passing throught the mixed phase.This feature is simply due to the fact that our hpdf is
directly connected to the density of states in each phase.
\newline
Let us now see what the plots of the normalized cumulants in Fig.(3-13) express? The general behavior and the structure of the peaks are much different. However the broadening effect of the transition region with decreasing volume is also observed.
The plots of skewness, kurtosis and pentosis, show a double peaks structure,
a big peak and a little one. These two peaks correspond to the two states
before and after the phase transition. When the two peaks have the same
sign, there are two vanishing points limiting the transition region and
containing a small extremum which is nothing other than the transition point. This behavior is due to the fact that kurtosis is closely connected the second derivative of the thermal susceptibility. Otherwise there is only one vanishing point which is the transition point. The
only difference between the three curves lies on the fact that the small
peak becomes less pronounced with increasing order of the cumulant. For
this reason, the latter does not appear practically on the curves.
In the transition region the symmetric peak of $p(\mathbf{h}=0.5,T)$ becomes very
small by making the kurtosis negative and small. The
kurtosis manifests a very different behavior in both sides of the transition
region when approaching the thermodynamic limit which is due to the
high asymmetry of the variance, as displayed clearly on the 3-Dim plot in Fig.(4). The variance
decreases more sharply in the hadronic phase than in the CQGP phase.
When looking more closely at all the 3-dimensional plots, we can clearly see that some particular points exhibit a typical behavior that can be described by the finite size scaling law, which is consistent with what has been obtained previously \cite{Ladrem2005}. For example, the maximum of the variance, sketches the finite size scaling behavior described by : $ T(\sigma _{\max })-T_{0}(\infty ) \propto {V}^{-1}$. Concerning the plots of the three hexosis, namely $\mathcal{H}_{1,2,3}$, a same global behavior out of the transition region and a different oscillatory behavior in it. The local maximum point in $\mathcal{H}_{1}$ becomes a singularity point in $\mathcal{H}_{2}$ and a local minimum in $\mathcal{H}_{3}$. Moreover, the obvious change in the sign, observed in our results, is in agreement with the results obtained by other models \cite{SignCumulant}.
Finally the plots in Fig.11,12,13 represent the variations of the three
 unnormalized cumulant ratios $\mathcal{O}\left( T,V\right) ,\mathcal{U}\left(
T,V\right) $ and $\mathcal{N}\left( T,V\right) $ as a function of
temperature and volume. Their behaviors are very different compared to the
plots of the normalized ratios. The plots of $\mathcal{N}\left( T,V\right) $
show a clear and rapid oscillatory behavior with two maxima and one
minimum in the transition region which gradually narrows as the volume increases.
On the other side we can clearly see the emergence of particular singular
behavior on the plots of $\mathcal{O}\left( T,V\right) $ and $\mathcal{U}%
\left( T,V\right)$ at certain values of temperature. The same divergence
is observed on the plot of the pentosis $\mathbf{\Pi }_{+}\left( T,V\right)$, exactly in the valley region between the two maximums (Fig.9). It is interesting to note the behaviour of $\mathcal{O}\left( T,V\right) $ which is practically zero in the two phases and is singular at the finite volume transition point, with a small local minimum before the transition and small local maximum after the transition.The location of the finite volume transition point is clear and simple, its shifting is obvious. The same observations are valid for $\mathcal{U}\left( T,V\right)$.
Using FSS analysis, we will see below that these points will be identified as the finite
volume transition points. We summarize by saying that $\mathcal{O}\left(
T,V\right) $ and $\mathcal{U}\left( T,V\right) $ tend to zero rapidly
everywhere, except in the transition region and at the finite volume transition point where they diverge.
This, is due to the zero of skewness in the
transition point. These two cumulant ratios can therefore serve as two good indicators of the location of the finite volume transition point. They will be of great use in the analysis of experimental data of URHIC where the context of initial conditions just before the phase transition are unknown.
We can see again from the figures that change their values sharply from negatives to positives and oscillate greatly with temperature near the transition point. These qualitative features; ie, sign change and oscillating structure, are consistent with effective models \cite{EffModels}.
\begin{figure}
\resizebox{0.50\textwidth}{!}{%
  \includegraphics{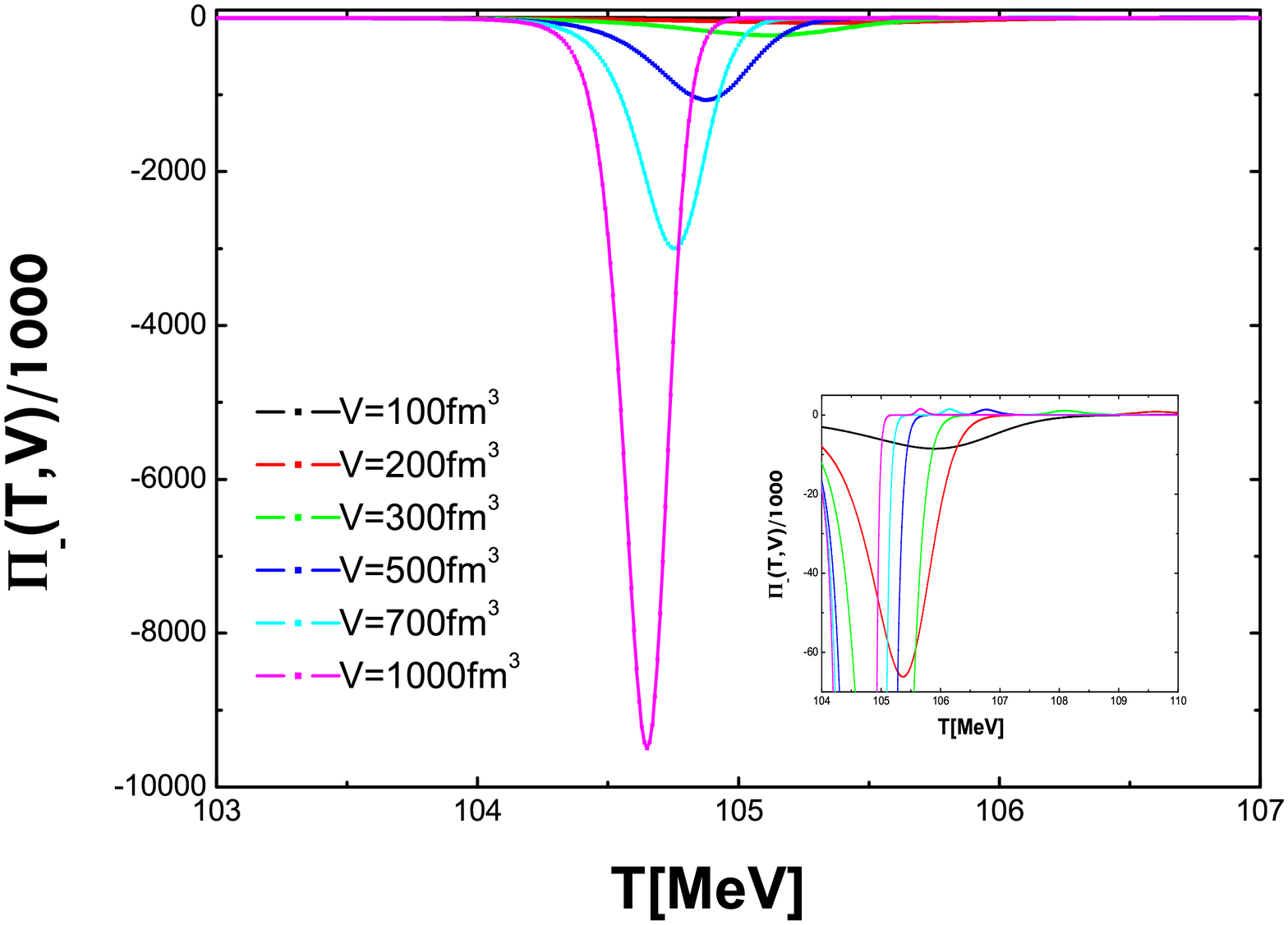}
}
\caption{Pentosis $\mathbf{\Pi }_{-}\left( T,V\right)$ vs Temperature for different Volumes(+ zoom of the region close to zero).}
\label{pen+2dim}       % Give a unique label
\end{figure}
\begin{figure}
\resizebox{0.50\textwidth}{!}{%
  \includegraphics{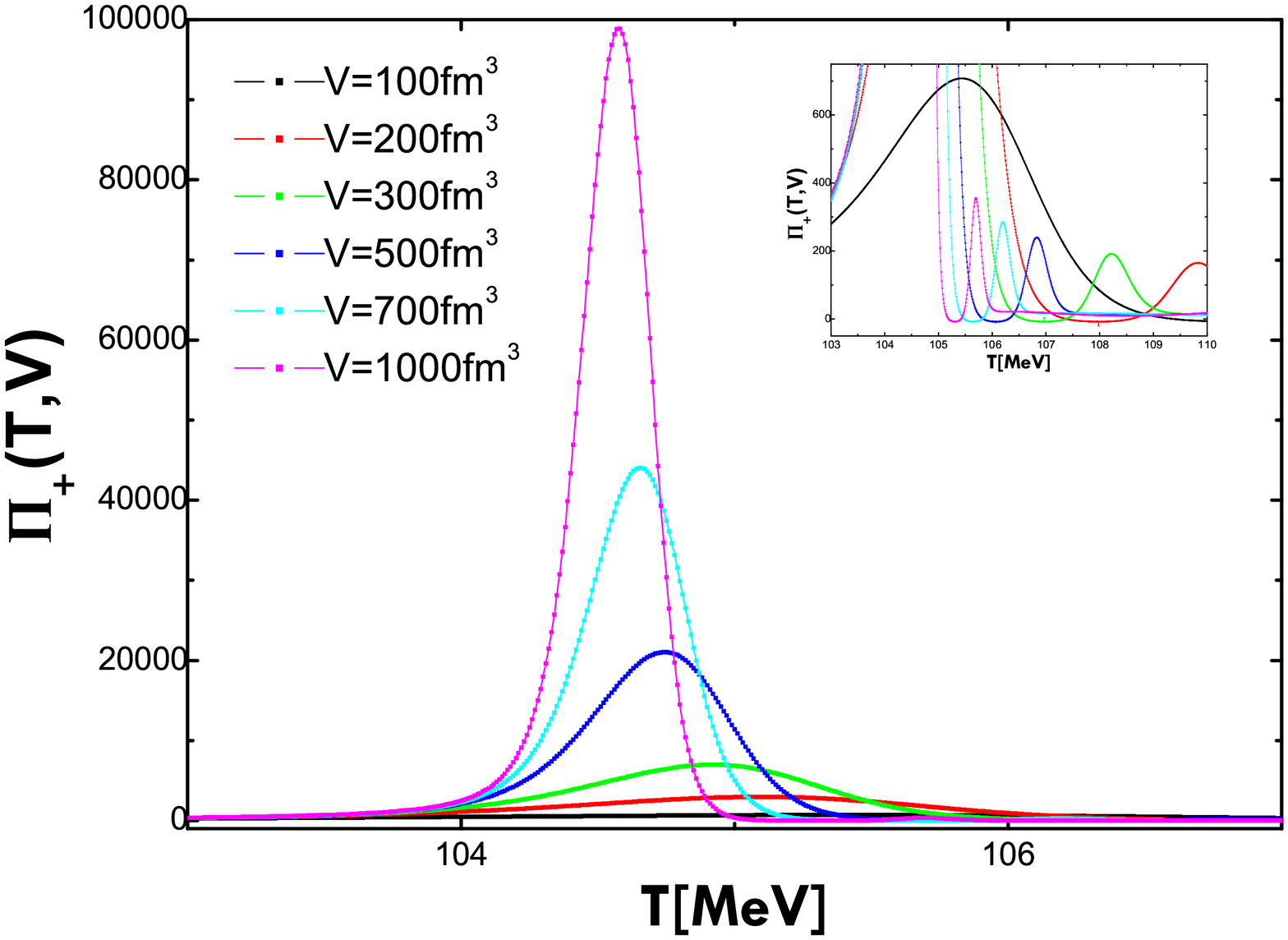}
}
\caption{Pentosis $\mathbf{\Pi }_{+}\left( T,V\right)$ vs Temperature for different Volumes (+ zoom of the region close to zero).}
\label{pen-2dim}       % Give a unique label
\end{figure}
\begin{figure}
\resizebox{0.50\textwidth}{!}{%
  \includegraphics{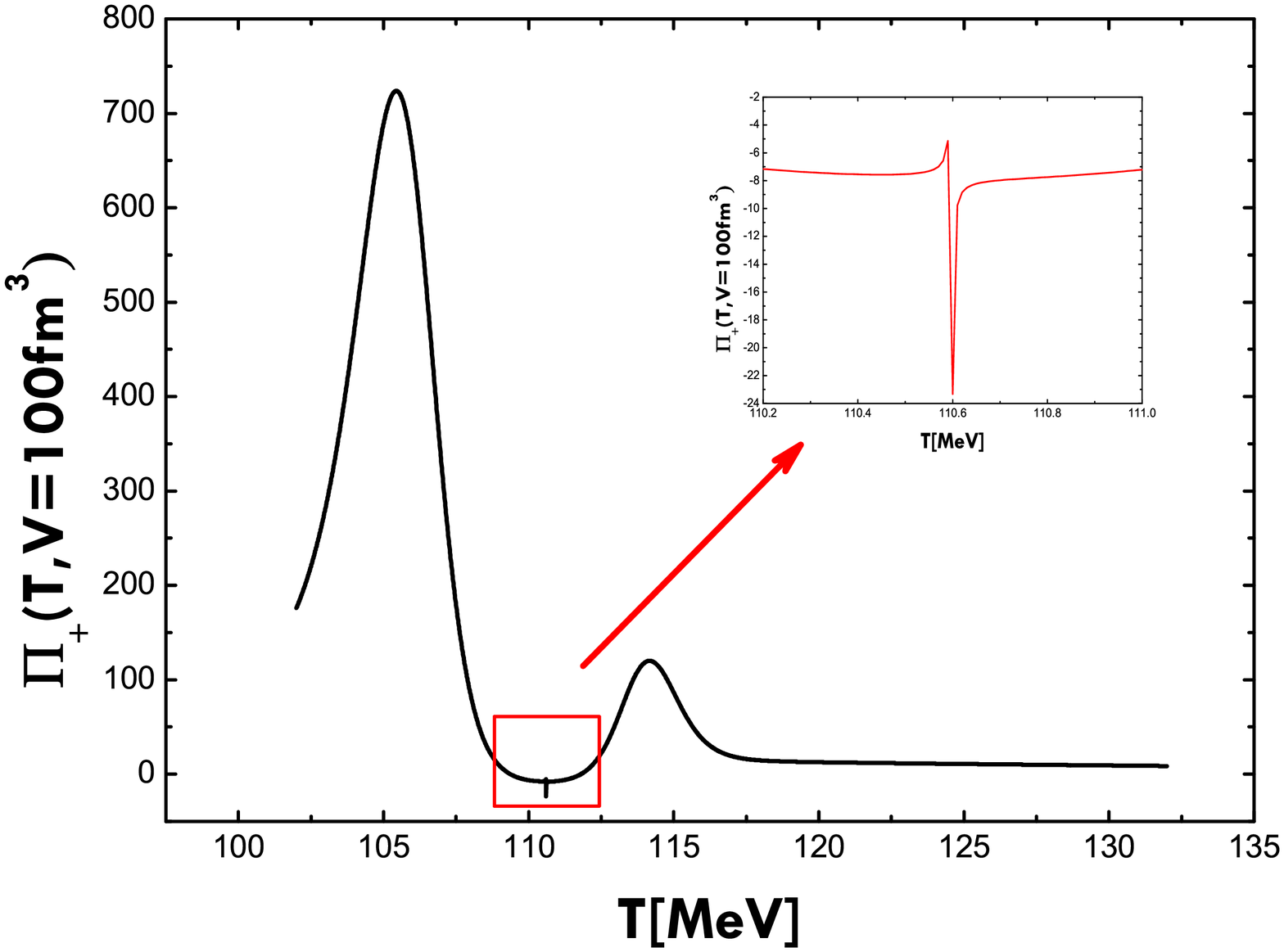}
}
\caption{Pentosis $\mathbf{\Pi }_{+}\left( T,V\right)$ vs Temperature for Volume $=100fm^{3}$.}
\label{pent100}       % Give a unique label
\end{figure}
\begin{figure}
\resizebox{0.50\textwidth}{!}{%
  \includegraphics{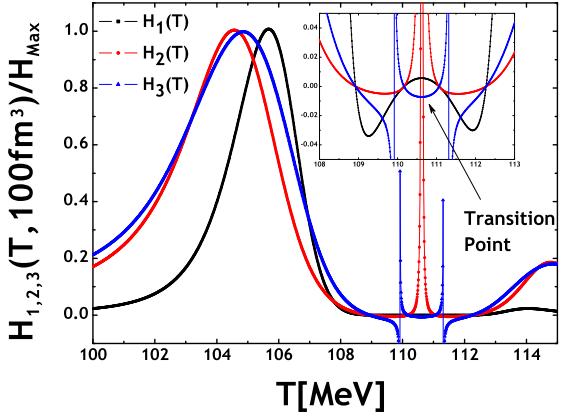}
}
\caption{Hexosis $\mathcal{H}_{1,2,3}\left( T,V\right)$ vs Temperature for Volume $=100fm^{3}$ (+ zoom of the region close to zero).}
\label{hex2dim}       % Give a unique label
\end{figure}
\begin{figure}
\resizebox{0.50\textwidth}{!}{%
  \includegraphics{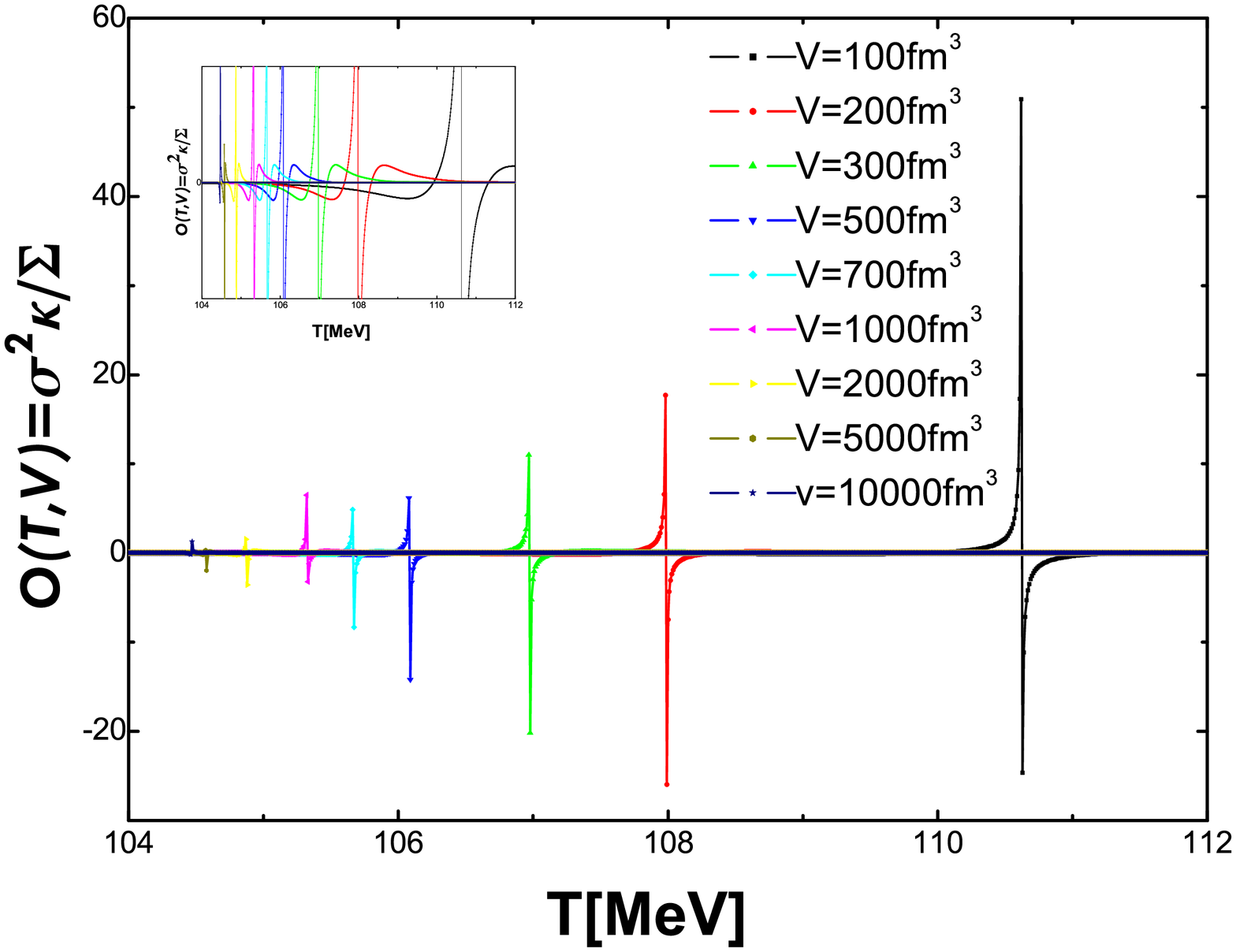}
}
\caption{$\mathcal{O}\left( T,V\right)$ vs Temperature for different Volumes (+ zoom of the region close to zero).}
\label{o2dim}       % Give a unique label
\resizebox{0.50\textwidth}{!}{%
  \includegraphics{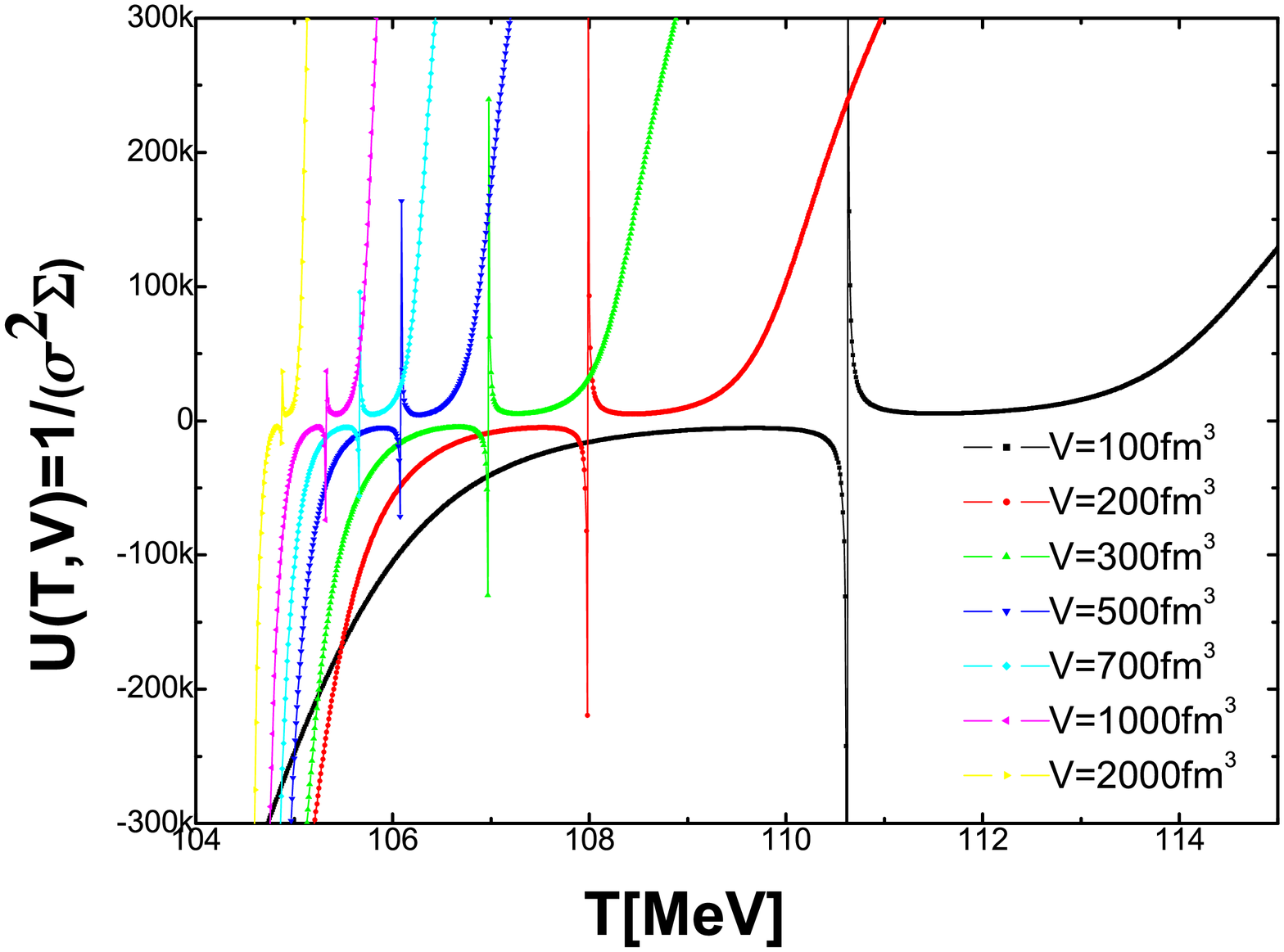}
}
\caption{$\mathcal{U}\left( T,V\right)$ vs Temperature for different Volumes.}
\label{u2dim}       % Give a unique label
\resizebox{0.50\textwidth}{!}{%
  \includegraphics{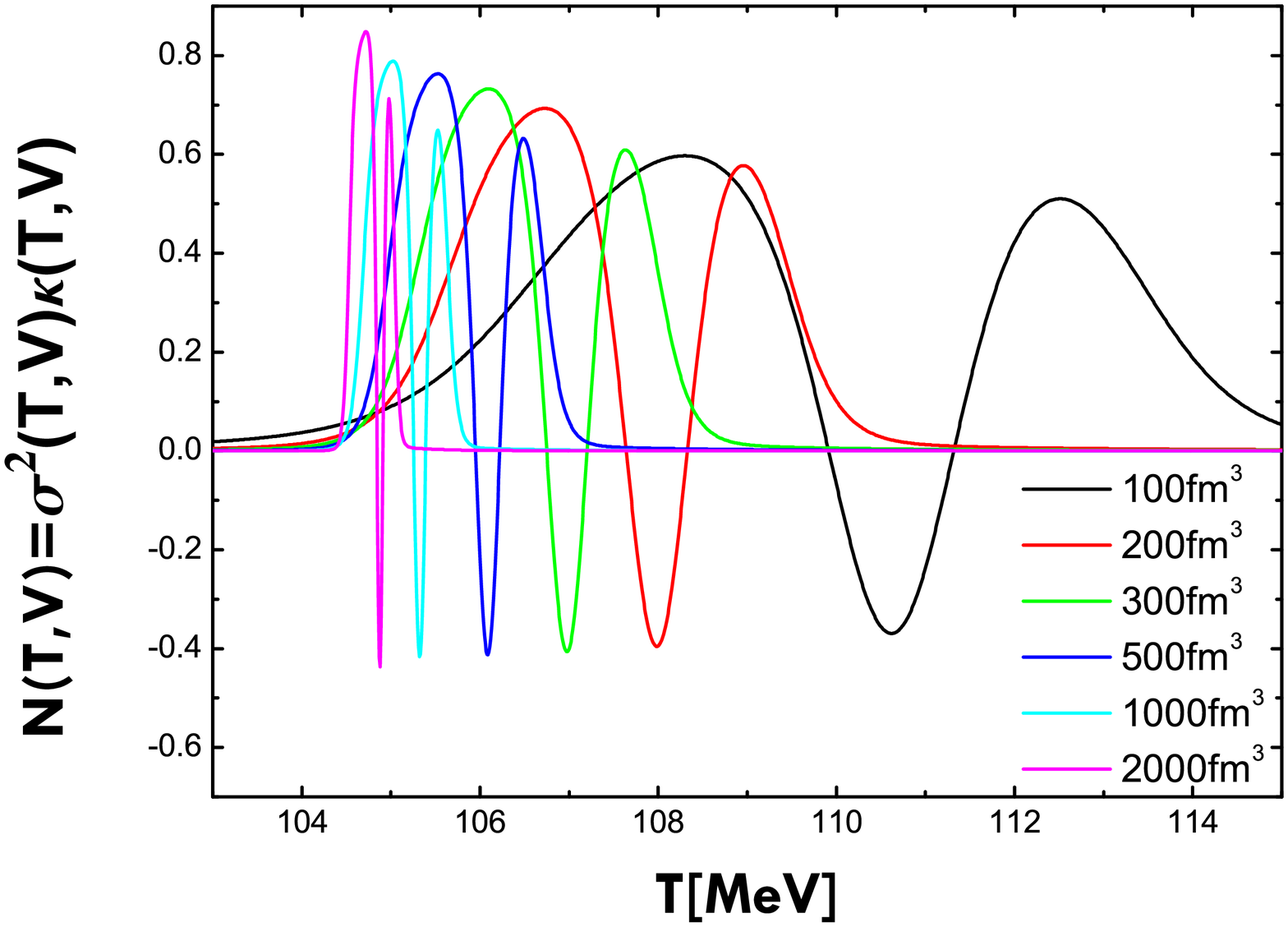}
}
\caption{$\mathcal{N}\left( T,V\right)$ vs Temperature for different Volumes.}
\label{n2dim}       % Give a unique label
\end{figure}
\section{New Method of Localization of the Finite Volume Transition Point}
\subsection{Natural Method}
It is important to have a precise knowledge of the region around the
transition point since many quantities of physical interest are just defined
in its vicinity. It therefore seems very important to
find the definition of a finite-volume transition point which involves less
corrections. Let us first remind the logical and natural way to define
the finite volume transition point by saying that is the point where we
have equal probabilities between hadronic phase and CQGP phase: $\left\langle
\mathbf{h}\left( T_{0}^{N}(V)\right) \right\rangle =1-\left\langle \mathbf{h}%
\left( T_{0}^{N}(V)\right) \right\rangle $. This means that the value of the order
parameter is given by $\langle \mathbf{h}\left( T_{0}^{N}(V)\right) \rangle
=1/2.$ We know that in thermodynamic limit the order parameter manifests a
finite discontinuity which can be easily described by a step function
(\ref{AsymThLim}). Therefore, the specific heat $%
c_{T}\left( T,V\right) $ and the thermal susceptibility $\chi _{T}\left(T,V\right) $ show $\delta $-function singularities at the transition point :
\begin{equation}
\lim_{(V)\to \infty}
\left\{
\begin{array}{c}
c_{T}\left( T,V\right)\\
\chi _{T}\left( T,V\right)%
\end{array}%
\right\}
\propto \delta(T-T_{0}(\infty )) \label{DelatcX}
\end{equation}
 In finite volume, these $\delta $-singularities become rounded peaks. Therefore $\chi
_{T}\left( T,V\right) $ and $c_{T}\left( T,V\right) $ reach a local extremum value at
certain temperature $T_{0}^{N}(V)$ which is defined as the temperature of the finite volume transition
point :
\begin{equation}
\left\{
\begin{array}{c}
c_{T}\left( T,V\right)=\textrm{max.}\\
\chi _{T}\left( T,V\right)=\textrm{min.}%
\end{array}%
\right\} \textrm{when $T=T_{0}^{N}(V)$} \label{ExtremumcX}
\end{equation}
Finally, we can assert without any problem that the finite volume transition point (and its temperature $T_{0}^{N}(V)$) is logically the point where the following equations are satisfied :
\begin{equation}
\left\{
\begin{array}{c}
\langle \mathbf{h}\left( T_{0}^{N}(V)\right) \rangle =1/2 \\
\left. \frac{\partial \chi _{T}\left( T,V\right) }{\partial T}\right\vert
_{T_{0}^{N}(V)}=0\ \left. \textrm{and  }  \frac{\partial c_{T}\left( T,V\right) }{%
\partial T}\right\vert _{T_{0}^{N}(V)}=0%
\end{array}%
\right. . \label{NaturalTP}
\end{equation}
From this we see that the finite volume transition point is associated to the appearance of an inflection point in $\langle \mathbf{h}(T,V)\rangle$: $\langle \mathbf{h}\left( T_{0}^{N}(V)\right) \rangle$ becoming a local extremum point in both  $\chi _{T}\left( T,V\right)$ and $c_{T}\left( T,V\right)$.
According to this method, we extract the different temperatures $T_{0}^{N}(V)$ of the transition points and collect them in table (\ref{tab:2}).
\subsection{Cumulant Method : Particular Points and Correlations}
In this section, we will try to propose a new method for locating the finite volume transition point using the whole cumulants studied in this work. We shall show how this finite volume transition, clearly manifests itself as a particular point in each cumulant.
\begin{table}
\caption{Natural Transition Points Temperatures}
\label{tab:2}       % Give a unique label
\begin{tabular}{|l|l|}
\hline
\textbf{Volume $\mathbf{V}$}$\left[ fm^{3}\right] $ & $T_{0}^{N}(V)\left[ MeV\right] $ \\ \hline
$100$ & $\mathbf{110.68007\pm 0.00001}$ \\ \hline
$200$ & $\mathbf{108.02068\pm 0.00001}$ \\ \hline
$300$ & $\mathbf{107.00271\pm 0.00001}$ \\ \hline
$400$ & $\mathbf{106.44758\pm 0.00001}$ \\ \hline
$500$ & $\mathbf{106.09471\pm 0.00001}$ \\ \hline
$700$ & $\mathbf{105.66892\pm 0.00001}$ \\ \hline
$900$ & $\mathbf{105.41823\pm 0.00001}$ \\ \hline
$1000$ & $\mathbf{105.32709\pm 0.00001}$ \\ \hline
$2000$ & $\mathbf{104.88739\pm 0.00001}$ \\ \hline
$5000$ & $\mathbf{104.59347\pm 0.00001}$ \\ \hline
$10000$ & $\mathbf{104.48254\pm 0.00001}$ \\ \hline
$\infty $ & $\mathbf{104.34796}$ \\ \hline
\end{tabular}%
\end{table}
Our strategy, we use, consists of finding a judicious point where the temperature $T_{0}(V)$, seemingly tends to the bulk $T_{0}(\infty )$\ with
increasing volume and must be highly correlated with $T_{0}^{N}(V)$ :
\begin{equation}
\lim_{(V)\to \infty}T_{0}(V)=T_{0}(\infty ).
\label{TLTemp}
\end{equation}
\begin{figure}
\resizebox{0.50\textwidth}{!}{%
  \includegraphics{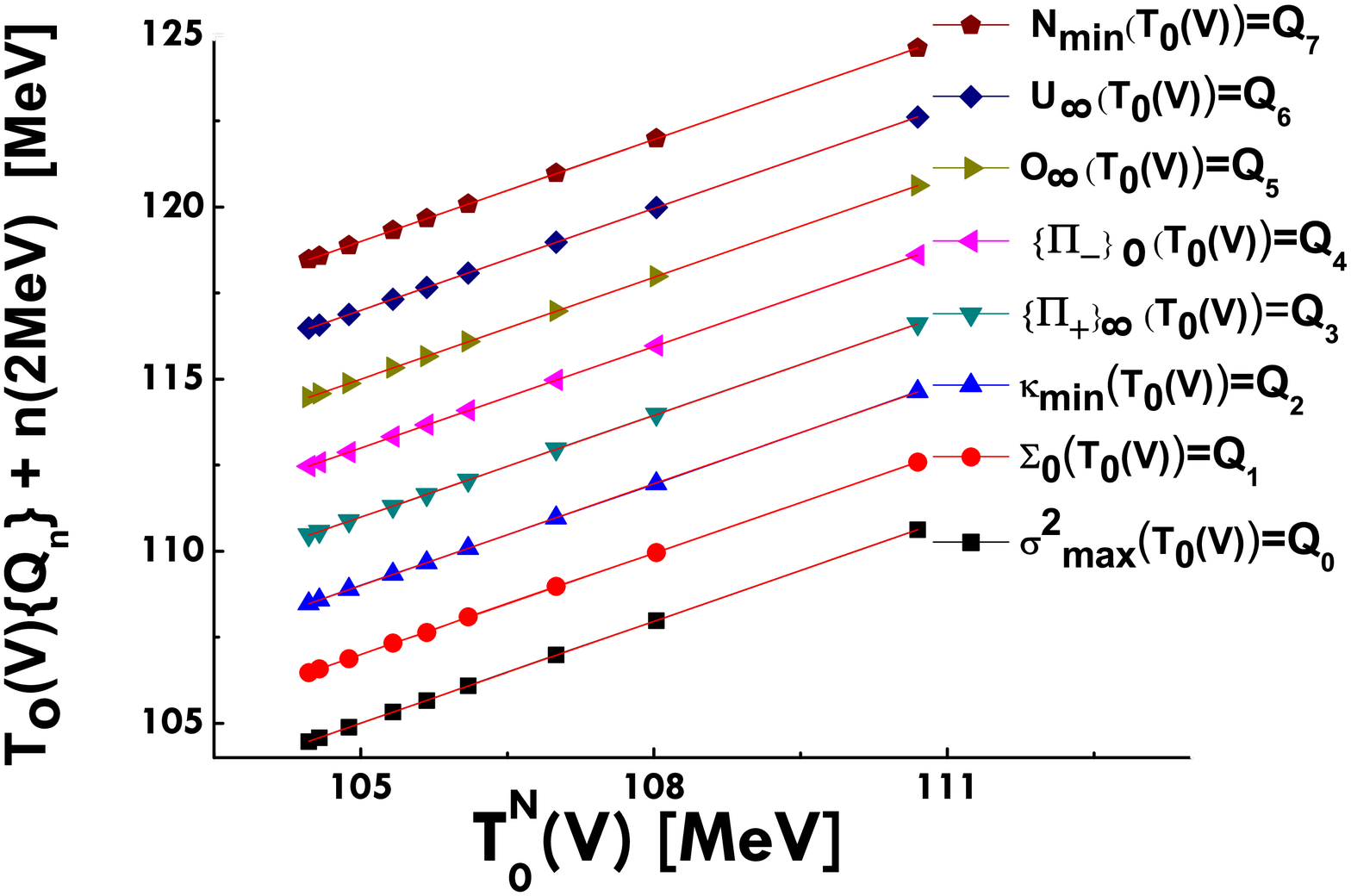}
}
\caption{Correlation scatter plot between $T_{0}(V)\lbrace {Q}_{n}\rbrace + n(2MeV) $ and $ T_{0}^{N}(V)$ for different volumes $({Q}_{0}={\sigma }^{2}_{max}, {Q}_{1}={\Sigma }_{0}, {Q}_{2}={\kappa }_{min }, {Q}_{3}=({\Pi }_{+})_{\infty }, {Q}_{4}=({\Pi }_{-})_0, {Q}_{5}={O}_{\infty }, {Q}_{6}={U}_{\infty}, {Q}_{7}=N _{min})$.}
\label{corrtot}       % Give a unique label
\end{figure}
The definition of $T_{0}(V)$ is not arbitrary but very difficult analytically and differs according to the quantity being considered. After a careful analysis of the normalized cumulants plots $\sigma
^{2}(T,V),\Sigma (T,V),\kappa (T,V),\Pi _{\pm}(T,V),$\\
$\mathcal{H}_{1,2,3}(T,V),\mathcal{O}(T,V),\mathcal{U}(T,V)$ and $\mathcal{N}(T,V)$, we find that the only
points which can be considered in one way or another as very particular are
: the local extrema points (local maximum and local minimum), the vanishing points (zeros), the inflection points and the singular points. These points are called the Particular Points. Indeed, we have investigated the behaviour of these particular
points. Firstly, for each quantity and for each particular point, we extract the temperature values $\left\{ T_{0}(V)\right\} $ at different volumes and put them in the first set.
Secondly we put the temperature values $\left\{ T_{0}^{N}(V)\right\} $ given in table (\ref{tab:2}) in the second set.
To probe more precisely the location of the finite volume transition point, a useful
tool is the scatter plot, in which the temperatures of the first set are
plotted against the temperatures of the second set. What we are asking here is whether or not the variations in the first set of $T_{0}(V)$ are correlated or not with the variations in the second set of $T_{0}^{N}(V)$.
We have analyzed several particular points and only good candidates are considered in this work with details. If a particular point is considered as a good finite volume transition point, one would expect that its scatter plot satisfies the following three criteria:
\newline
(1) The fit should be linear.
\newline
(2) The slope of the fit should equal unity and its vertical intercept should equal zero.
\newline
(3) The fit should have high linear correlation with a very good correlation factor and a very good probability test.
\newline
If we consider the temperature $\left\{ T_{0}(V)\right\} $ to be dependent variable, then we want to know if the scatter plot can be described by a linear function of the form,
\begin{equation}
T_{0}(V)={\lambda} T_{0}^{N}(V)+{\nu}.  \label{Correlation}
\end{equation}
Because we are discussing the relationship between the variables $\left\{ T_{0}(V)\right\} $ and $\left\{ T_{0}^{N}(V)\right\}$, we can also consider $\left\{ T_{0}^{N}(V)\right\} $ as a function of $\left\{ T_{0}(V)\right\} $ and ask if the data follow the same linear behavior,
\begin{equation}
T_{0}^{N}(V)={\lambda'} T_{0}(V)+{\nu'}.  \label{Correlation2}
\end{equation}
The values of the coefficients ${\lambda'}$ and ${\nu'}$ in (\ref{Correlation2}) will be different from the values of the coefficients  ${\lambda}$ and ${\nu}$ in equation (\ref{Correlation}), but they are related if the two temperatures $\left\{ T_{0}(V)\right\} $ and $\left\{ T_{0}^{N}(V)\right\}$ are correlated.
If we consider solely the value of $\lambda$ (or $\lambda'$), it doe not provide us a good measure of the degree of the correlation.
From (\ref{Correlation}) and (\ref{Correlation2}), and in the case of a total correlation, we can show that
\begin{equation}
\left\{
\begin{array}{c}
{\lambda\lambda'}=1 \\
\lambda\nu'+\nu=0
\end{array}%
\right. .  \label{TotaleCorrelation}
\end{equation}
If there is no correlation, the two parameters ${\lambda}$ and ${\lambda'}$ are lower than unity, even approaching zero value. We therefore can use the product ${\lambda\lambda'}$ as a measure of the correlation between the two sets of temperatures $\left\{ T_{0}(V)\right\} $ and $\left\{ T_{0}^{N}(V)\right\}$. By definition the correlation factor is given by ${\varrho\equiv\sqrt{\lambda\lambda'}}$.
The value of ${\varrho}$ ranges from $0$ , when the data are totally uncorrelated, to $1$, when there is total correlation.
The correlation factor, alone, is not sufficient to indicate the quality or the goodness of the linear fit. An additional calculation of probability is necessary for more precision. This probability distribution enables us to go beyond the simple fit, and  to compute a probability associated with it. In the case of our situation, a commonly used probability distribution for ${\varrho}$  is given by \cite{PughWinslow1966,Fornasini2008},
\begin{equation}
\mathcal{P}_{\varrho}(\varrho,\zeta)=\frac{1}{\sqrt{\pi }}\frac{\Gamma[(\zeta+1)/2]}{\Gamma[(\zeta)/2]}(1-{\varrho}^2)^{(\zeta-2)/2}, \label{CorrProbability}
\end{equation}
where $\zeta=N-2$ is the number of degrees of freedom for a sample of $N$ data points, and $\Gamma(x)$ is the standard Gamma function.It gives the probability that any sample of uncorrelated data would yield to a linear behavior described by a correlation factor equal to ${\varrho}$.  If this probability is small, then the sample of data points can be considered as highly correlated variables. More generally, this type of calculation is often referred to as goodness of fit test \cite{Bevington2003}.
Another significant and useful quantity which can be calculated from the distribution(\ref{CorrProbability}) is given by,
\begin{equation}
\mathcal{P}_{C}(\varrho,N)=2\int\limits_{|\varrho|}^{1} \mathcal{P}_{x}(x,\zeta)d{x}.
\end{equation}
This $P_{C}(\varrho,N)$  represents the integral probability that a sample of $N$ uncorrelated data points would yield a linear correlation factor larger or equal than the calculated value of $|\varrho|$.
This would mean that a small value of $P_{C}(\varrho,N)$ is equivalent to a high probability that the two sets of variables are linearly correlated.
The fitting results obtained from the correlations study showed on Fig.(14), are summarized in table (\ref{tab:3}). In order to avoid overlapping between fitting curves and to allow a clear representation on the same graph, we have added a shift of 2 MeV between each two consecutive curves.
\begin{table}
\caption{Correlation factor values obtained from linear fitting}
\label{tab:3}       % Give a unique label
\begin{tabular}{|l|l|l|l|}
\hline
\textbf{N.Cumulant} & \textbf{Transition Point}& \textbf {$\lambda$}& \textbf {$\lambda'$}%
\\ \hline
$\sigma ^{2}(T,V)$ & $\sigma _{\max }^{2}(T_{0}(V))$
& $\mathbf{0.98812}$ & $\mathbf{1.01202}$\\ \hline
$\ \Sigma (T,V)$ &  $\ \Sigma _{0}(T_{0}(V))$ & $%
\mathbf{0.98798}$ & $\mathbf{1.01216}$\\ \hline
$\kappa (T,V)$ &  $\kappa _{\min
}(T_{0}(V))$ & $\mathbf{0.98700}$ & $\mathbf{1.01317}$ \\ \hline
$\Pi _{+}(T,V)$ & $\left( \Pi
_{+}\right) _{\infty }(T_{0}(V))$ & $\mathbf{0.98788}$ & $\mathbf{1.01226}$\\ \hline
$\Pi _{-}(T,V)$ &  $\left( \Pi _{-}\right)
_{0}(T_{0}(V))$ & $\mathbf{0.98753}$ & $\mathbf{1.01262}$\\ \hline
$\mathcal{O}(T,V)$ &  $\mathcal{O}_{\infty }(T_{0}(V))$ & $%
\mathbf{0.98787}$ & $\mathbf{1.01227}$\\ \hline
$\mathcal{U}(T,V)$ &  $\mathcal{U}_{\infty }(T_{0}(V))$ & $%
\mathbf{0.98787}$ & $\mathbf{1.01227}$\\ \hline
$\mathcal{N}(T,V)$ & $\mathcal{N}_{\min }(T_{0}(V))$ & $%
\mathbf{0.98753}$ & $\mathbf{1.01262}$\\ \hline
\end{tabular}%
\vspace*{0.2cm}  % with the correct table height
\end{table}
It can be perceived from the scatter plots Fig.14 that the points are closely scattered about an
underlying straight line, refelecting a strong linear relationship between the two sets of data and the numerical values of the slopes are close to unity as expected. Also, we tried the fitting procedure with a fixed intercept $\nu=0$ \ and we got better results, the value of the slope better than $0.999$.
From the values of both $\lambda$ and $\lambda'$ in the table 3, pratically a same value of the correlation factor $\varrho$, which equal to $0.99999$, is obtained. Therefore the evaluation of the two probabilities gives the following results:
\begin{equation}
\left\{
\begin{array}{c}
\mathcal{P}_{\varrho}(\varrho=0.99999,\zeta=7)=1.82209\times10^{-12} \\
\mathcal{P}_{C}(\varrho=0.99999,N=9)=1.04119\times10^{-17}
\end{array}%
\right. .  \label{Probabilities}
\end{equation}
The extreme smallness of $P_{C}(\varrho,N)\leq1.178\times10^{-16}$ indicates that it is extremely improbable that the variables under consideration are linearly uncorrelated. Thus the probability is very high that the variables are correlated and the linear fit is justified.
The fact that such fittings yield results that are consistent with each other is an important
consistency check on the accuracy of the calculations and gives an idea of the FSE for the values of the temperature of
finite volume transition point . We would like to note that the numerical values of temperature obtained by the cumulant method  $\left\{ T_{0}(V)\right\} $ of the various transition points, are comparable with an accuracy less than $2\%$, with the temperatures $\left\{ T_{0}^{N}(V)\right\} $ extracted using conventional procedures.
Therefore the selected points are indeed the true finite volume transition points, namely:
 \newline
(1) the local maximum point in the variance $\sigma ^{2}(T,V)$ and in the first hexosis $\mathcal{H}_{1}(T,V)$: $\sigma _{\max }^{2},\mathcal{H}_{1,max}$,
\newline
(2) the zero point in the skewness $\Sigma(T,V)$ and in the pentosis $\Pi _{-}(T,V)$: $\Sigma _{0}, \Pi _{-,0}$,
\newline
(3) the local minimum point in the kurtosis $\kappa(T,V)$, in $\mathcal{N}(T,V)$ and in the third hexosis $\mathcal{H}_{3}(T,V)$: $\kappa _{\min },\mathcal{H}_{3,min},$
$\mathcal{N}_{\min }$
\newline
(4) and the singularity point in the pentosis $\Pi _{+}(T,V)$, in $ \mathcal{U}(T,V)$, in $\mathcal{O}(T,V)$ and in the second hexosis $\mathcal{H}_{2}(T,V)$: $\Pi _{+,\infty},\mathcal{U}_{\infty }$, $\mathcal{O}_{\infty },\mathcal{H}_{2,\infty}$.
\newline
The temperature at which skewness vanishes is expected to represent the transition temperature, and tends apparently to $T_{0}(\infty )$\ with increasing volume, while the temperature gap between the two extrema is expected to give the width of the transition region.
\newline
We got an unexpected and important result. It concerns the behavior of the connected Binder cumulant. Indeed from the relation(34) the whole discussion about the kurtosis can be translated to the connected Binder cumulant. Therefore, the connected Binder cumulant $\mathcal{B}^{c}_{4}(T,V)$ has two minima and a little maximum between them as expected from the behavior of the kurtosis $\kappa (T,V)$. The position of two minima should not have a good correlation factor, however, the little maximum will be the good finite volume transition point. This would be in striking contrast to conventional result obtained by Binder \cite{CLB1986}. The apparent discrepancy is completely due to the difference in the defintion of the Binder cumulant $\mathcal{B}_{4}(T,V)$ and the connected Binder cumulant $\mathcal{B}^{c}_{4}(T,V)$. The local minimum point in the Binder cumulant is not the true finite volume transition point because it has not the good correlation factor ($\lambda=1.39$). But it should approach the bulk transition temperature as $V$ becomes large, which means that it is just a particular point. We have therefore shown that the cumulants are more interesting than the moments and the connected Binder cumulant is more efficient in locating the true finite volume transition point than the Binder cumulant. The same results are obtained in many papers \cite{Borgs1992,Velonakis2015,Janke1993} and the obtained thermal behaviors are in complete agreement with ours.
We know that all the particular points as they have been defined in our paper converge towards the unique singularity in the thermodynamic limit. Once the true finite volume transition point has been identified from the particular points, its signal is not necessarily the highest, and even, may be in some cases, is hard to detect. The main property of the particular points in finite volume is that they are correlated with the true finite volume transition point. Another important property relates to the possibility of using them to define a transition region.
It has been claimed in that the shift between the minimum of the Binder cumulant and the maximum in its susceptibility in the case of a first order phase transition, is due to the absence of the phase coexistence phenomena in the double Gaussian model and of the surface corrections \cite{Bhanot1989,Lee1991}. In our case, despite the intake into account of the phase coexistence within the Colorless-MIT bag model, the shift between the minimum of the Binder cumulant and the true finite volume transition point still exists but its magnitude is different. The magnitude of this shift is reflected in the numerical values of the correlation parameters $(\lambda,\nu)$ which differ from the ideal values $(\lambda=1,\nu=0)$ in the case of a total correlation. Indeed, when we try to extract roughly the numerical values of $\lambda$ parameter from the results obtained in \cite{CLB1986,Lee1991,Janke1993,Martinos2005}, we find different values [$\lambda=1.55,1.47,1.57,1.89$] respectively, which are not close to unity. This is certainly due to the fact that our Colorless-MIT bag model is very different from the double gaussian model used by Binder to study the finite size effects in the first order phase transition \cite{CLB1986}. Presumably the shift of the minimum of $\mathcal{B}_{4}(T,V)$ from the true finite volume transition point $T_{0}^{N}(V)$ depends on the detailed form of the partition function of the system under consideration as quoted in \cite{Lee1991}, ie, it is somewhere model-dependent.
\section{Conclusion}
In order to identify and locate the finite volume transition point more accurately, we have studied in details the finite volume cumulant expansion of the order parameter and have shown how greatly this can be used to provide a clear definition of the finite volume transition point in the context of the thermal deconfinement phase transition to a CQGP.
Starting from the hadronic probability density function and using the $L_{mn}$-method, a finite size cumulant expansion of the order parameter is carried out. The first six cumulants, their under-normalized ratios and also some combinations of them, are then calculated and analyzed as a function of temperature at different volumes. To be more consistent and coherent in our definitions of cumulant ratios, a new reformulation of these cumulant ratios is proposed.
It has been put into evidence that all cumulants and their ratios showed deviations from their asymptotic values(low and high temperature values), which increase with the cumulant order. This behavior is essential to discriminate the phase transition by measuring the fluctuations.
We have noticed that both cumulants of higher order and their ratios, associated to the thermodynamical fluctuations of the order parameter, in QCD behave in a particular enough way revealing pronounced oscillations in the transition region. The sign structure and the oscillatory behavior of these in the vicinity of the deconfinement phase transition point might be a sensitive probe and may allow to elucidate their relation to the QCD phase transition point.
In the context of our model, we have shown that the finite volume transition point is always associated to the appearance of a particular point in whole cumulants under consideration.
A detailed FSS analysis of the results has allowed us to locate the finite volume transition points and extract accurate values of their temperatures $T_{0}(V)$. We have tested the validity of our results by performing linear correlations between the set of $T_{0}(V)$ and the known results obtained with the natural definition $T_{0}^{N}(V)$ providing very good correlation factors.
In addition to natural definition of the finite volume transition point as the extrema of thermal susceptibility, $\chi _{T}$ and specific heat $c_{T}$, we have shown that the true finite volume transition point manifests itself as a different particular point according to the quantity considered, namely as,
\newline
(1) a local maximum point in the variance $\sigma ^{2}(T,V)$ and in the first hexosis $\mathcal{H}_{1}(T,V)$: $\sigma _{\max }^{2},\mathcal{H}_{1,max}$,
\newline
(2) a zero point in the skewness $\Sigma(T,V)$ and in the pentosis $\Pi _{-}(T,V)$: $\Sigma _{0}, \Pi _{-,0}$,
\newline
(3) a local minimum point in the kurtosis $\kappa(T,V)$, in $\mathcal{N}(T,V)$ and in the third hexosis $\mathcal{H}_{3}(T,V)$: $\kappa _{\min },\mathcal{H}_{3,min},$
$\mathcal{N}_{\min }$
\newline
(4) a singularity point in the pentosis $\Pi _{+}(T,V)$, in $ \mathcal{U}(T,V)$, in $\mathcal{O}(T,V)$ and in the second hexosis $\mathcal{H}_{2}(T,V)$: $\Pi _{+,\infty},\mathcal{U}_{\infty }$,
\newline
$\mathcal{O}_{\infty },\mathcal{H}_{2,\infty}$.
\newline
It is important to mention that the finite volume transition point, using the connected Binder cumulant $\mathcal{B}^{c}_{4}(T,V)$, is given by the little maximum $\left( \mathcal{B}^{c}_{4}\right) _{\max }$ between the two minima. By against, the minimum of the Binder cumulant $\mathcal{B}_{4}(T,V)$ : $\left( \mathcal{B}_{4}\right) _{\min }$ as obtained in \cite{CLB1986,Borgs1992,Velonakis2014,Velonakis2015,Janke1993,Martinos2005}, is just a particular point and not the true finite volume transition point. Obviously any particular point tends to the bulk transition point as $V$  becomes large. The apparent discrepancy is completely due to the difference in the defintion of the Binder cumulant $\mathcal{B}_{4}(T,V)$ and the connected Binder cumulant $\mathcal{B}^{c}_{4}(T,V)$. The shift between $\left( \mathcal{B}_{4}\right) _{\min }$ and the true finite volume transition point in our model is different to those obtained by other models. This is probably due to the fact that our hpdf is very different from the double gaussian distribution used by Binder \cite{CLB1986} and that considered in \cite{Lee1991}. We therefore suspect that this shift is somewhere model dependent as quoted in \cite{Lee1991}. We will present a detailed study of this point in a forthcoming work. Finally, we can conclude that the finite volume transition point that appears as a particular point, the emergence of the linear correlation between different particular points, and the possibility to use them to define a transition region, are the features of a \textit{universal} behavior.
\newpage
\
%\newpage
\section{appendix}
From the general expression of the mean value $\left\langle \mathbf{h}^{n}\right\rangle (T,V)$ (\ref{meanhnLmn}), we can easily deduce the first eight mean values :
\begin{equation}
\langle \mathbf{h}\rangle (T,V)=\frac{L_{02}\left( 1\right) -L_{02}\left(
0\right) -L_{01}\left( 0\right) }{L_{01}\left( 1\right) -L_{01}\left(
0\right) },  \label{OPLmn}
\end{equation}%
\begin{equation}
\langle \mathbf{h}^{2}\rangle (T,V)=\frac{2L_{03}\left( 1\right)
-2L_{03}\left( 0\right) -2L_{02}\left( 0\right) -L_{01}(0)}{L_{01}\left(
1\right) -L_{01}\left( 0\right) },  \label{MVh2}
\end{equation}%
\begin{equation}
\langle \mathbf{h}^{3}\rangle (T,V)=
\frac{6L_{04}\left( 1\right)
-6L_{04}\left( 0\right) -6L_{03}\left( 0\right) -3L_{02}\left( 0\right)
-L_{01}(0)}{L_{01}\left( 1\right) -L_{01}\left( 0\right) },  \label{MVh3}
\end{equation}
\begin{equation}
\langle \mathbf{h}^{4}\rangle (T,V)=\frac{24L_{05}\left( 1\right)
-24L_{05}\left( 0\right) -24L_{04}\left( 0\right) -12L_{03}\left( 0\right)
-4L_{02}\left( 0\right) -L_{01}(0)}{L_{01}\left( 1\right) -L_{01}\left(
0\right) }.  \label{MVh4}
\end{equation}
\begin{equation}
\langle \mathbf{h}^{5}\rangle (T,V)=\frac{120L_{06}\left( 1\right)
-120L_{06}\left( 0\right) -120L_{05}\left( 0\right) -60L_{04}\left( 0\right)
-20L_{03}\left( 0\right) -5L_{02}\left( 0\right) -L_{01}(0)}{L_{01}\left(
1\right) -L_{01}\left( 0\right) }.  \label{MVh5}
\end{equation}
\begin{equation}
\langle \mathbf{h}^{6}\rangle (T,V)=\frac{720L_{07}\left( 1\right)
-720L_{07}\left( 0\right) -720L_{06}\left( 0\right) -360L_{05}\left(
0\right) -120L_{04}\left( 0\right) -30L_{03}\left( 0\right)
-6L_{02}(0)-L_{01}(0)}{L_{01}\left( 1\right) -L_{01}\left( 0\right) }.
\label{NVh6}
\end{equation}
\begin{equation}
\langle \mathbf{h}^{7}\rangle (T,V)=\frac{5040L_{08}\left( 1\right)
-5040L_{08}\left( 0\right) -5040L_{07}\left( 0\right) -2520L_{06}\left(
0\right) -840L_{05}\left( 0\right) -210L_{04}\left( 0\right)
-42L_{03}(0)-7L_{02}(0)-L_{01}(0)}{L_{01}\left( 1\right) -L_{01}\left( 0\right) }.
\label{NVh7}
\end{equation}
\begin{multline}
\langle \mathbf{h}^{8}\rangle (T,V)=\frac{40320L_{09}\left( 1\right)
-40320L_{09}\left( 0\right) -40320L_{08}\left( 0\right) -20160L_{07}\left(
0\right) -6720L_{06}\left( 0\right) -1680L_{05}\left( 0\right)
-336L_{04}(0)}{L_{01}\left( 1\right) -L_{01}\left( 0\right) } \vspace*{0.25cm} \\
-\frac{56L_{03}(0)+8L_{02}(0)+L_{01}(0)}{L_{01}\left( 1\right) -L_{01}\left( 0\right) }.
\label{NVh8}
\end{multline}
\clearpage
\section{acknowledgments}
This research work was supported in part by the Deanship of Scientific
Research at Taibah University (Al-Madinah, KSA) under Contract 432/765 and
also by the King Abdulaziz City for Science and Technology under contract
No. (P-S-12-0660).M.L. would like to dedicate this work in living memory of
his daughter Ouzna Ladrem (violette) died suddenly in March 24,2010. May
Allah has mercy on her and greet her in his vast paradise. Many thanks to
A. Y. Jaber from M.L and M.A.A.A. for his infinite availability and
great support during their stay at Al-Madinah.


\end{document}